\documentclass[lettersize,journal]{IEEEtran}
\usepackage{amsmath,amsfonts}
\usepackage{algorithmic}
\usepackage{array}
\usepackage{bm}
\usepackage{bbm}
\usepackage[caption=false,font=normalsize,labelfont=sf,textfont=sf]{subfig}
\usepackage{textcomp}
\usepackage{stfloats}
\usepackage{braket}
\usepackage{threeparttable}
\usepackage{tablefootnote}
\usepackage{diagbox}
\usepackage{url}
\usepackage{hyperref}
\usepackage{verbatim}
\usepackage{multicol}
\usepackage{multirow}
\usepackage{bbding}
\usepackage[numbers,longnamesfirst]{natbib}
\usepackage{graphicx}
\def\BibTeX{{\rm B\kern-.05em{\sc i\kern-.025em b}\kern-.08em
    T\kern-.1667em\lower.7ex\hbox{E}\kern-.125emX}}
\usepackage{balance}
\begin{document}
\title{Quantum Circuit Synthesis and Compilation Optimization: Overview and Prospects}
\author{Ge Yan \textit{Member, IEEE}, Wenjie Wu, Yuheng Chen, Kaisen Pan, Xudong Lu,\\ Zixiang Zhou, Yuhan Wang, Ruocheng Wang, Junchi Yan \textit{Senior Member, IEEE }
\thanks{The authors are with the MoE Key Lab of Artificial Intelligence, Shanghai Jiao Tong University, Shanghai 200240, China (email: yanjunchi@sjtu.edu.cn). Work was partly supported by NSFC (72342023).}}

\markboth{Quantum Circuit Synthesis and Compilation Optimization: Overview and Prospects}{}


\maketitle

\begin{abstract}
Quantum computing is a promising paradigm that may overcome the current computational power bottlenecks. The increasing maturity of quantum processors provides more possibilities for the development and implementation of quantum algorithms. As the crucial stages for quantum algorithm implementation, the logic circuit design and quantum compiling have also received significant attention, which covers key technologies, e.g., quantum logic circuit synthesis (also widely known as quantum architecture search) and optimization, as well as qubit mapping and routing. Recent studies suggest that the scale and precision of related algorithms are steadily increasing, especially with the integration of artificial intelligence methods. In this survey, we systematically review and summarize a vast body of literature, exploring the feasibility of an integrated design and optimization scheme that spans from the algorithmic level to quantum hardware, combining the steps of logic circuit design and compilation optimization. Leveraging the exceptional cognitive and learning capabilities of AI algorithms, it becomes more possible to reduce manual design costs, enhance the precision and efficiency of execution, and facilitate the implementation and validation of the superiority of quantum algorithms on hardware. An actively maintained paper list is available at: \href{https://github.com/Thinklab-SJTU/awesome-ml4Qcircuit}{https://github.com/Thinklab-SJTU/awesome-ml4Qcircuit}.
\end{abstract}

\begin{IEEEkeywords}
quantum circuit synthesis, quantum compiling, quantum machine learning, artificial intelligence
\end{IEEEkeywords}

\section{Introduction}
\IEEEPARstart{I}{n} recent years, quantum computing and quantum machine learning have garnered extensive attention. Due to their unique physical properties, quantum computing is considered the most promising new computing architecture to break through the existing computational power bottlenecks in the post-Moore era. Several notable quantum algorithms, e.g. Shor's algorithm for large integer factorization~\cite{shor1994algorithms} and Grover's algorithm for unstructured search~\cite{grover1996fast}, have been theoretically proven to enjoy quantum superiority over classical algorithms. With the advent of the Noisy Intermediate-Scale Quantum (NISQ) era in the past decade~\cite{huang2020superconducting,preskill2018quantum}, quantum superiority has not only been demonstrated theoretically but also experimentally verified on quantum prototypes. For example, Google has demonstrated quantum superiority on a superconducting quantum\footnote{For clarity and conciseness, all references to quantum hardware in this article refer to superconducting systems, if not otherwise specified. } prototype through Random Circuit Sampling (RCS) tasks~\cite{arute2019quantum,wu2021strong}. Similarly, the team with USTC has verified this through Gaussian Boson Sampling (GBS) on a photonic quantum prototype~\cite{zhong2020quantum}.

Exploring the superiority of non-trivial and practically significant quantum algorithms beyond RCS and GBS presents a significant challenge, particularly given the constraints of current quantum hardware, including limited coherence time, noise, and physical topology. Many quantum algorithms rely on oracles~\cite{shor1994algorithms,grover1996fast,weinstein2001implementation}, which are black-box unitary matrices with known operations but undefined implementations. Implementing these algorithms requires efficient realization of these unitary operators, minimizing key metrics such as circuit depth and gate count. The corresponding logic circuits are then optimized based on specific goals. In the NISQ era, the primary focus has been on optimizing two-qubit gates, while in the fault-tolerant quantum computing (FTQC) era, optimization targets, such as T gates, become more prominent. Once optimized, logic circuits must undergo compilation, which involves mapping logical qubits to physical qubits and inserting necessary SWAP gates to meet connectivity constraints. For variational quantum algorithms (VQAs), specific synthesis and optimization are needed for parameterized circuits. Balancing hardware constraints with algorithmic goals, we aim to optimize circuit performance by minimizing noise while maintaining theoretical precision, recognizing the trade-off between circuit depth and execution errors.

\begin{figure*}[tb!]
    \centering
    \includegraphics[width=\textwidth]{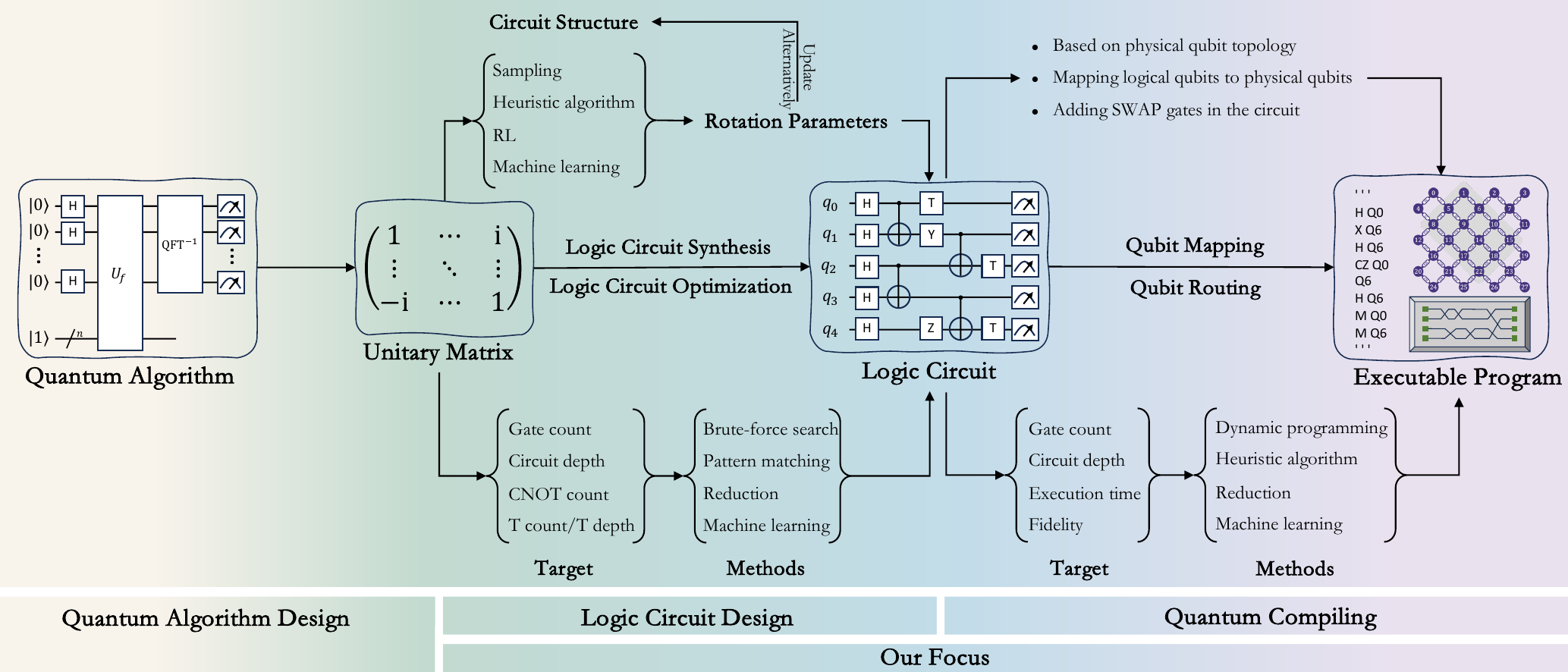}
    \vspace{-17pt}
    \caption{Quantum algorithm implementation pipeline. A quantum algorithm can be written in the form of multiple unitary transformations. logic circuit synthesis and optimization methods are then applied to obtain logic circuits. We utilize qubit mapping and routing methods to build an executable program during the quantum compiling stage. In this paper, we mainly focus on logic circuit design and quantum compiling.}
    \label{fig:overall}
\end{figure*}

Fig.~\ref{fig:overall} illustrates the entire process from quantum algorithm design to execution. This process includes transforming a quantum algorithm into unitary transformations, generating logic circuits through synthesis and optimization methods, and then compiling these circuits, considering the physical qubit topology and other quantum hardware constraints, into executable quantum programs on target quantum processors. In other words, the transformation from a quantum algorithm to an executable quantum program can be divided into three steps: quantum algorithm design, quantum logic circuit design, and quantum compiling. In reviews of modern quantum computing technology~\cite{huang2023near}, these steps are often collectively referred to as quantum circuit compilation, with relevant research briefly summarized. Early work also roughly summarized quantum circuit compilation in the NISQ era with heuristic methods~\cite{kusyk2021survey}, including genetic algorithms, genetic programming, ant colony optimization, and planning algorithms.
In this paper, we focus on efficiently transforming quantum algorithms into executable quantum programs by comprehensively summarizing research on logic circuit synthesis and optimization, as well as qubit mapping and routing. Additionally, we provide prospects and insights into automatic quantum logic circuit design and compilation optimization, which enables researchers to focus more on algorithm design, fully leveraging the potential of NISQ-era quantum hardware and promoting the practical application and superiority verification of quantum algorithms.

This survey introduces methods for representing and optimizing quantum circuits, covering quantum circuit representations such as gate models~\cite{deutsch1985quantum}, directed acyclic graphs~\cite{maslov2008quantum}, phase polynomials~\cite{amy2014polynomial}, ZX diagrams~\cite{coecke2008interacting}, and tensor networks~\cite{biamonte2017tensor}. We explore synthesis and optimization methods tailored to these representations.
For quantum logic circuit synthesis, we discuss applications with and without target unitary matrices. Applications with given unitary matrices include oracle implementation~\cite{bijwe2022implementing,rahman2022grover}, while those without target unitary matrices include VQAs~\cite{grimsley2019adaptive,du2022quantum,wu2023quantumdarts}, error correction codes~\cite{chen2019machine,cong2019quantum}, and nonlinear unit design~\cite{rattew2023non,guo2024nonlinear}. Quantum circuit synthesis typically involves determining circuit structure and evolving rotation gate parameters, with machine learning increasingly improving optimization over traditional methods.
For quantum circuit optimization, we categorize relevant literature based on objectives and methods. Quantum circuit optimization improves circuits based on optimization goals and is influenced by the different characteristics of NISQ and FTQC eras. Methods include pattern matching, peephole optimization~\cite{mckeeman1965peephole}, and reductions to well-studied problems~\cite{heyfron2018efficient,amy2019t,schneider2023sat}. Recent advances in reinforcement learning~\cite{fosel2021quantum,li2023quarl,riu2023reinforcement} have further improved optimization by discovering new matching patterns, surpassing traditional approaches. Optimization methods for VQAs, such as those for specific VQAs~\cite{grimsley2019adaptive,tang2021qubit,yordanov2021qubit,magoulas2023linear}, enhance their suitability for NISQ-era quantum processors.
In quantum circuit compilation, we address physical qubit connectivity and the need for SWAP gates to handle limited connectivity between qubits. While SWAP gates enable qubit state exchanges, they introduce overhead in terms of gate count and can impact operational accuracy. Therefore, quantum qubit mapping and routing algorithms are essential for optimizing gate counts, circuit depth, runtime, and fidelity~\cite{tannu2019not,murali2019noise,niu2020hardware,liu2022not,liu2021qucloud,niu2023enabling}. These methods, including exact solutions~\cite{siraichi2018qubit}, heuristics~\cite{zulehner2018efficient}, reductions~\cite{murali2019noise,wille2019mapping,molavi2022qubit}, and machine learning approaches~\cite{pozzi2022using,sinha2022qubit,huang2022reinforcement}, focus on optimizing the performance of quantum circuits.

This leads to a key question explored in this paper: is it possible to develop an integrated algorithm that encompasses both quantum logical circuit design and quantum compiling? Such an algorithm should include the following steps:
(1) Generate the optimal logical circuit corresponding to the quantum algorithm based on the built-in gate set of the quantum processor;
(2) Complete the selection, mapping, and routing of quantum bits and couplers according to the calibration data of the quantum processor;
(3) Balance theoretical accuracy and practical noise, achieving the optimal performance of the quantum processor with an appropriate trade-off in theoretical accuracy. Solving all these constraints is a challenging problem. However, the introduction of artificial intelligence may facilitate the design of related algorithms. Relevant literature indicates that with the introduction of machine learning algorithms, quantum logical circuit synthesis, optimization, and compilation have become significant application scenarios for AI4Science. Ideally, integrating all the above steps into the compiler of the quantum hardware cloud platform could automate these processes through AI algorithms. Ideally, integrating these steps into a quantum hardware cloud platform’s compiler, powered by AI algorithms, could automate the process, significantly improving the practical performance of submitted algorithms and eliminating the need for manual intervention after algorithm design.

\textbf{In summary, the contributions of this paper are as follows:} (1) Conducting a comprehensive analysis of the various steps and challenges in the development and execution of quantum algorithms, covering the entire process from quantum logical circuit design to quantum compiling; (2) Systematically organizing the algorithms and their characteristics related to all sub-tasks, providing abstract mathematical definitions for each step, facilitating the introduction of AI methods and lowering the entry barrier for newcomers; (3) Analyzing the requirements for real-world implementation of quantum algorithms in the NISQ era, and proposing an integrated quantum circuit design and compiling solution. This solution leverages artificial intelligence to enhance the performance and design efficiency of quantum algorithms and envisions a new paradigm for quantum compiling in the age of AI.

\begin{figure*}[tb!]
\centering
    \includegraphics[width=\textwidth]{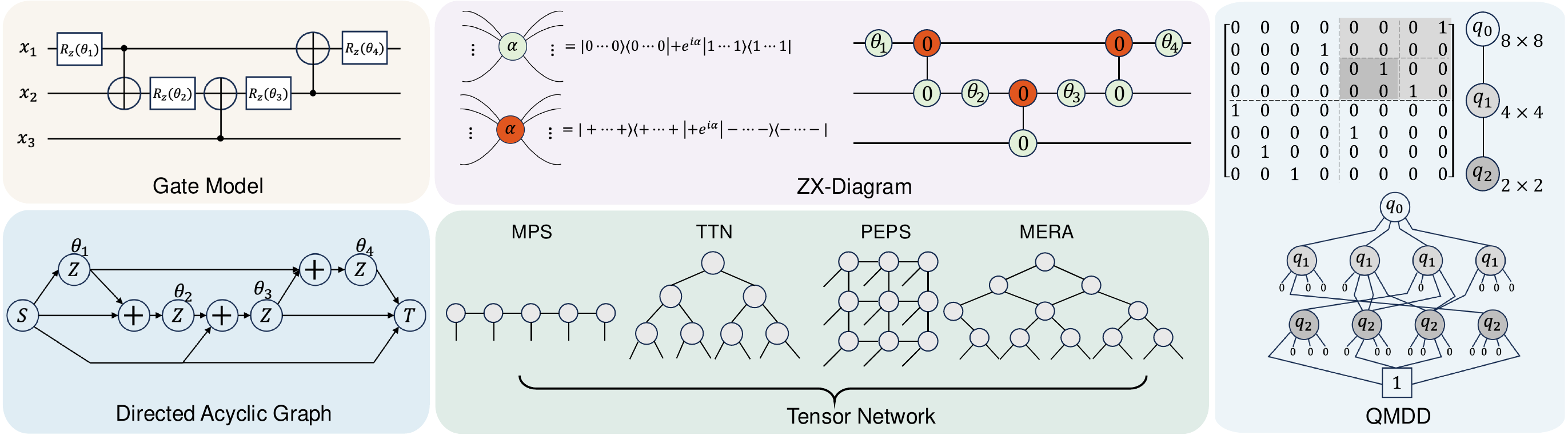}\vspace{-7pt}
    \caption{Quantum Circuit Representation. Gate Model: a popular representation method; DAG: example DAG transformed from the gate model example, ``start'' and ``end'' nodes are added to complete the graph; QMDD: QMDD representation of a three-qubit quantum operation; we present the transformation matrix, the layerwise structure and the size of the corresponding matrix of each node; Tensor Network: a quantum state $\ket{\psi}$ can be represented by multiple common types of tensor network; ZX-Diagram: the green and red spider in ZX-diagram as well as an example ZX-diagram transformed from gate model example.}
    \label{fig:circuit_representation}
\end{figure*}

\section{Quantum Circuit Representation}
\label{sec:circuit_rep}
We begin by discussing the quantum circuit representations. This involves using graphs or vectors to depict specific circuit information, crucial for their synthesis and optimization. Different synthesis and optimization algorithms are designed based on particular representations. Choosing the right representation can reduce the complexity of subsequent design and compiling steps while still conveying the necessary information.

\subsection{Quantum Gate Model}
The most common and fundamental method for representing quantum circuits is through the quantum gate model. This method characterizes quantum circuits using a system similar to classical gate circuits by representing the operations on qubits as quantum gates. A quantum gate can operate on multiple qubits. Each gate corresponds to a unitary transformation, modifying the quantum state and enabling circuit operations.
Quantum gates can be divided into two main categories: parameterized and non-parameterized gates. The former allows for continuous transformations. To implement quantum algorithms, a set of quantum gates that can approximate any unitary matrix with arbitrary precision is required. This set is known as the universal quantum gate set~\cite{nielsen_chuang_2010}. According to~\cite{nielsen_chuang_2010}, the necessary and sufficient condition for a universal quantum gate set is that it can approximate any single-qubit unitary transformation and includes a two-qubit gate. 

One of the most commonly used and well-studied universal quantum gate sets is Clifford+T~\cite{forest2015exact}. The Clifford set includes H$=\frac{1}{\sqrt{2}}\begin{bmatrix}
1&1\\1&-1
\end{bmatrix}$, S$=\begin{bmatrix}
1&0\\0&e^{\text{i}\frac{\pi}{2}}
\end{bmatrix}=\sqrt{\text{Z}}$, and CNOT
gates~\cite{gottesman1997stabilizer,gottesman1998theory}. Quantum circuits containing only Clifford set can be efficiently simulated by classical computers, making this set particularly important~\cite{gottesman1998heisenberg}. 

Through the combination of these three gates, we can generate a variety of other gates. For example, the Pauli-Z gate can be obtained by S$^2$, the Pauli-X gate can be obtained by HZH$^\dagger$, and the Pauli-Y gate can be obtained by SXS$^\dagger$. However, these basic Clifford gates are not universal on their own. They cannot approximate many other gates with arbitrary precision within a finite sequence. To achieve universal quantum computation, we need to supplement the Clifford set with the T gate: $\text{T}=
\begin{bmatrix}
1&0\\
0&e^{\text{i}\frac{\pi}{4}}
\end{bmatrix}=\sqrt{\rm S}=\sqrt[4]{\rm Z}$.
The T gate is a phase gate. When we supplement the Clifford set with the non-Clifford T gate, the quantum gate set can continuously generate new non-Clifford gates, making the {H, S, T, CNOT} set capable of approximating any quantum gate. The Clifford+T set is the most commonly used universal quantum gate set, primarily because Clifford gates are relatively easy to implement on superconducting quantum computers. However, the T gates are more challenging to implement, so one major objective of quantum circuit optimization is to minimize the number of T gates required.

Besides Clifford+T, there are parameterized universal gate sets. Consider the parameterized rotation gates $\rm Rx(\theta)=\begin{bmatrix}
\cos(\frac{\theta}{2})&-\text{i}\sin(\frac{\theta}{2})\\-\text{i}\sin(\frac{\theta}{2})&\cos(\frac{\theta}{2})
\end{bmatrix}$, $\rm Ry(\theta)=\begin{bmatrix}
\cos(\frac{\theta}{2})&-\sin(\frac{\theta}{2})\\\sin(\frac{\theta}{2})&\cos(\frac{\theta}{2})
\end{bmatrix}$, and $\rm Rz(\theta)=\begin{bmatrix}
e^{-\text{i}\frac{\theta}{2}}&0\\0&e^{\text{i}\frac{\theta}{2}}.
\end{bmatrix}$. Along with the CNOT gate, they form a commonly used gate set in quantum machine learning. 
Compared to the Clifford+T set, the main advantage of this set is that all single-qubit gates are parameterized, allowing for continuous control over rotation angles. As a result, the number of gates required to approximate arbitrary single-qubit gates is significantly reduced compared to using Clifford+T gates with discrete angles. However, their physical implementation on hardware is limited by the resolution of the control electronics. This leads to inherent errors in gate operation, as the analog relationship between pulse duration and rotation angle cannot be realized with infinite precision.

Additionally, some universal gate sets are designed for specific problems. For instance, the NCV quantum gate set is used to optimize circuits~\cite{miller2011elementary,maslov2003simplification, maslov2005quantum,maslov2008quantum}, particularly multi-controlled-NOT gates. This set includes NOT gates, CNOT gates, and Controlled-V/V$^\dagger$ gates, where the unitary matrix $\rm V=\frac{1+\text{i}}{2}\begin{bmatrix}1&-\text{i}\\-\text{i}&1\end{bmatrix}$. 
Other studies directly utilize hardware built-in gates when synthesizing and optimizing circuits, e.g., iSWAP$=e^{\text{i}\frac{\pi}{4}(XX+YY)}$~\cite{stanisic2022observing}. These special cases will be explained in detail when encountered in the subsequent sections.

\subsection{Directed Acyclic Graph}
To better extract information from quantum circuits, we can use Directed Acyclic Graphs (DAGs) to represent quantum gate circuits and apply them to tasks, e.g., quantum architecture search~\cite{he2023gsqas,he2023gnn} and quantum compiling~\cite{li2019tackling,beaudoin2024altgraph}. By adding a source node and a sink node at the beginning and end of the quantum circuit, respectively, and treating each quantum gate as a node, we create a directed graph. Adjacent gates on the same qubit are connected by directed edges, and multi-qubit gates introduce additional edges. These directed edges represent the sequence in which the gates are applied.
The advantages of using DAGs to represent quantum circuits are: (1) the structure of quantum circuits naturally transforms into DAGs; (2) DAGs can represent the application order of quantum gates and the control qubits of multi-qubit gates, maximizing the advantages of DAGs
; and (3) graph neural networks can effectively extract and compute features of nodes and edges, handling large-scale quantum circuits.

There is a special circuit representation in the form of DAGs called Quantum Multiple-Valued Decision Diagram (QMDD)~\cite{niemann2015qmdds}. As shown in Fig.~\ref{fig:circuit_representation}, starting from the $q_2$ level, i.e. a $2\times 2$ matrix, we build the diagram layer by layer according to the transformation matrix. Each $2\times 2$ matrix consists of four elements. There are four types of $q_2$ level matrices in the circuit, so this level has four nodes. The $q_1$ level above comprises a $4\times 4$ matrix. Each $q_1$ level matrix is composed of four $q_2$ level matrices, connecting each $q_1$ level node to four $q_2$ level nodes. The final $q_0$ level $8\times 8$ unitary matrix is the target unitary matrix we aim to build. The completed QMDD diagram is shown at the bottom of the QMDD module.

\subsection{Circuit Polynomial}
Based on the representation of the quantum gate model, we can further characterize quantum circuits using algebraic methods, e.g. linear Boolean functions or phase polynomials~\cite{amy2013meet,nam2018automated,meuli2018sat,amy2018controlled,vandaele2022phase}. These polynomial representation algorithms focus more on the quantum state rather than the specific quantum gates acting on the state. An arbitrary single-qubit quantum gate can be seen as inducing the following transformations on the quantum state:
\begin{equation}
\ket{x}\rightarrow e^{\text{i}p(x)}\ket{g(x)},
\end{equation}
where $\ket{x}\rightarrow \ket{g(x)}$ is a reversible transformation, and $e^{\text{i}p(x)}$ 
represents the phase information imparted by the quantum gate to the current quantum state $\ket{x}$. Since reversible transformations in quantum circuits are relatively straightforward to analyze, phase polynomial expressions focus more on the phase information contributed by quantum gates, particularly gates like Z, S, T, Rz$(\theta)$ and CZ. Given that the T gate is a crucial target in quantum circuit optimization, phase polynomial expressions play a significant role in optimizing circuits. Below, we present a specific example of a phase polynomial. For an $n$-qubit circuit, assuming that there are only two types of gates, CNOT and Rz$(\theta)$, the phase function $p(x)$ for the sample quantum circuit in Fig.~\ref{fig:circuit_representation} is:
\begin{equation}\label{eq:phasepoly}
\begin{aligned}
    p(x_1,x_2,x_3)&=\theta_1(x_1)+\theta_2(x_1\oplus x_2)\\&+\theta_3(x_1\oplus x_2\oplus x_3)+\theta_4(x_2\oplus x_3).
\end{aligned}
\end{equation}
The quantum states before and after the execution of the transformation circuit can be expressed as:
\begin{equation}
    \ket{x_1,x_2,x_3}\rightarrow e^{\text{i}p(x_1,x_2,x_3)}\ket{x_2\oplus x_3,x_1\oplus x_2\oplus x_3,x_3}.
\end{equation}
Since the coefficients in the phase function are all rotation angles limited to a range of $2\pi$, we can take the modulo of the superimposed angle. This allows the phase function to serve as a criterion for determining whether two circuits are equivalent. Compared to directly comparing two unitary matrices, using phase polynomials enables the equivalence assessment of larger-scale circuits. For instance, \cite{ruiz2025quantum} optimized a 72-qubit quantum circuit based on phase polynomial, a task that cannot be accomplished by merely using unitary matrices.

\subsection{Tensor Networks}
A tensor~\cite{penrose1971applications} can be regarded as a multidimensional vector, where scalars, vectors, and matrices can be considered as 0-, 1-, and 2-dimensional tensors, respectively~\cite{orus2014practical}. \cite{deutsch1985quantum} introduced the graphical representation of tensors into quantum computing, leading to the quantum gate circuits we discussed earlier. Therefore, the quantum gate model is indeed a special form of tensor networks. We will briefly introduce this more fundamental representation, with more detailed content about the tensor network available in~\cite{biamonte2017tensor}. Tensor networks can connect different tensors through indices, an operation called tensor network contraction. Some classic tensor network structures include Matrix Product State (MPS), Projected Entangled Pair States (PEPS), Tree Tensor Network (TTN), and Multi-Scale Entanglement Renormalization Ansatz (MERA) -- see Fig.~\ref{fig:circuit_representation}.

Since matrix multiplication can be viewed as the contraction of two second-order tensors based on the same index, quantum circuits can naturally be simulated and optimized using tensor networks~\cite{markov2008simulating}. The main optimization method is to reduce the cost of contractions, but finding an optimal tensor network contraction sequence is NP-hard~\cite{chi1997optimizing}. This problem can be further modeled as finding an optimal tree decomposition of the tensor network graph structure~\cite{schutski2020simple}. The tree pruning method proposed in the paper can decompose the original graph into multiple parallel subgraphs with the minimum tree depth. \cite{huang2021efficient} proposed index slicing, which divides the entire contraction into sub-tasks that can be executed in parallel. Specifically, a greedy algorithm selects indices that lead to the greatest reduction in complexity, accompanied by local reordering of the contraction tree. Additionally, pre-computing small tensors on the CPU and reordering matrix multiplication on the GPU can reduce the actual running cost. Contraction tasks can be divided into several tensor transpositions and one matrix multiplication~\cite{lyakh2015efficient}. \cite{vincent2022jet} constructed a task graph based on the dependencies of the aforementioned transpositions and multiplications, allowing different transpositions to be parallelized. Furthermore, by naming to ensure that shared operations between sliced contractions are only calculated once, costs are significantly reduced.

Another research direction is to optimize performance by simulating various tensor network structures when designing quantum circuits. These circuits are often referred to as Tensor Network Quantum Circuits (TNQC). Generative and adversarial tasks are completed using quantum circuits based on MPS and TTN~\cite{huggins2019towards}. Since TNQC allows us to simulate certain quantum circuits efficiently, we can use this method to save physical qubits and reduce the impact of real quantum computing noise. 
Quantum circuits based on MPS and MERA are also analyzed in~\cite{haghshenas2022variational}, finding that TNQC can use fewer quantum gate parameters to achieve the same accuracy, further validating the practicality of TNQC.

\subsection{ZX Diagrams}
ZX diagrams are a method for representing quantum circuits based on ZX calculus, proposed in~\cite{coecke2008interacting}. ZX diagrams are also a special form of tensor networks that can represent any linear mapping between qubits~\cite{Kissinger_2020}. They use spiders and the lines connecting spiders to represent a quantum circuit. Typically, ZX diagrams include two types of spiders: green Z spiders and red X spiders, as shown in Fig.~\ref{fig:circuit_representation}~\cite{yeung2020diagrammatic}. When the angles are $\pi$, these two types of spiders correspond to the Pauli-Z and Pauli-X gates, respectively.

Spiders define the basic linear operations in any dimension, where the lines on the left side of a spider represent the inputs of the current linear operation, and the lines on the right side represent the outputs. Each spider’s output can further connect to the inputs of other spiders, constructing a large network. ZX diagrams follow a set of fusion rules, which allow for obtaining many equivalent representations, simplifying the connections of complex spiders. Since 2019, much work has been done to optimize quantum circuits based on ZX diagrams. For instance,~\cite{Duncan_2020,kissinger2015quantomatic,Fagan_2019,van_de_Wetering_2021} studied the optimization of the number of spiders in ZX diagrams;~\cite{de_Beaudrap_2020,heyfron2018efficient,Munson_2021,Kissinger_2020_2} researched the optimization of T gate counts; and papers like~\cite{gogioso2022annealing,kissinger2019cnot} introduced heuristic algorithms for the contraction and optimization of ZX diagrams. By optimizing the ZX diagram representation of quantum circuits, we can obtain more streamlined quantum circuits. When these ZX diagrams are mapped back to gate circuits, we can get circuits that meet the requirements. Compared to directly optimizing quantum gate circuits, the optimization methods based on ZX diagrams can introduce different characteristics, making them a relatively common method for representing quantum circuits.

\section{Quantum Logic Circuit Synthesis}
In this section, we address the challenge of quantum logic circuit synthesis. Running a quantum algorithm on a quantum processor necessitates translating the algorithm's unitary transformations into quantum circuits. Typically, these circuits are designed manually~\cite{jaques2020implementing} or through decomposition algorithms~\cite{kitaev2002classical,vartiainen2004efficient,dawson2005solovay}. This process demands substantial human labor and expertise, often resulting in circuits that are particularly deep.
In the current NISQ era, various constraints must be considered, including the accuracy of single and double qubit gates, qubit decoherence time, and the topological structure of physical qubits. Constructing a quantum circuit that achieves the highest possible accuracy while adhering to these constraints is a pressing challenge. Therefore, this section delves into methods for automatically generating quantum circuits that meet these requirements. This automated synthesis of quantum logic circuits is also referred to as the Quantum Architecture Search (QAS) problem in literature~\cite {zhang2022differentiable,du2022quantum,lu2023qas}.

\subsection{Problem Definition}
First, we provide the mathematical definition of the QAS problem. The QAS problem originates from the Neural Architecture Search (NAS) problem, which automates the synthesis and optimization of neural network structures and parameters to replace human expert design. NAS has been widely applied in the field of computer vision, e.g. image classification~\cite{liu2018darts} and object detection~\cite{WangECCV22}, and has achieved success on multiple datasets. Similar to the NAS setting, when defining the QAS problem, we first determine the set of candidate gates and the search space. Assuming we need to search for a circuit with $n$ qubits and $m$ layers, define the set of candidate gates as $\mathcal{G}$, and the size of the candidate set as $k=|\mathcal{G}|$, then the search space of the QAS problem is of the size $k^{n \times m}$. The candidate gate set $\mathcal{G}$ can include single-qubit, two-qubit, and parameterized quantum gates. We denote the selected candidate gates as matrix $\mathbf{M}\in \mathcal{G}^{n \times m}$, define $\mathbf{U}_{ij}=\sigma(\mathbf{M}_{ij})$, where $\sigma$ maps a quantum gate on the $i$-th qubit and $j$-th layer to a $2^n \times 2^n$ unitary matrix with all the irrelevant qubits as identity. The unitary matrix corresponding to the searched circuit is:
\begin{equation}
\hat{\mathbf{U}}=\prod_{j=1}^m\prod_{i=1}^n\hat{\mathbf{U}}_{ij}=\prod_{j=1}^m\prod_{i=1}^n \sigma(\mathbf{M}_{ij}).
\end{equation}
For different applications, the QAS problem can be further divided into two versions. The first searches the circuit with a target unitary matrix $\mathbf{U}$ and the corresponding objective is:
\begin{equation}\label{eq:unitary}
\min ||\mathbf{U}-\hat{\mathbf{U}}||_F,
\end{equation}
where $||\cdot||_F$ represents the F-norm of the matrix.
The second one designs the circuit based on a series of quantum initial states $\ket{\chi_i}$ and corresponding final states $\ket{\psi_i}$, without an explicit unitary matrix. The corresponding objective is:
\begin{equation}\label{eq:non-unitary}
\max \sum_i |\bra{\psi_i}\hat{\mathbf{U}}\ket{\chi_i}|^2.
\end{equation}
Both versions of QAS problems have extensive and important applications, which we will introduce next.

\subsection{Applications of Quantum Architecture Search}
\subsubsection{Quantum Architecture Search with Unitary Matrix}
The most important application for QAS with unitary matrix is the circuit implementation of oracles. Here, an oracle refers to a black-box unitary matrix often seen in quantum algorithms. We may not know its specific implementation yet, but we understand the unitary operation corresponding to this oracle. Quantum oracle is of fundamental significance in quantum computing and quantum information, but implementing an arbitrary quantum oracle requires exponentially many gates~\cite{nielsen_chuang_2010}. Taking the oracle $\mathcal{O}$ in Grover's algorithm~\cite{grover1996fast} as an example, which can be represented as a unitary operation:
\begin{equation}
\ket{x}\ket{q}\stackrel{\mathcal{O}}{\rightarrow}\ket{x}\ket{q\oplus f(x)},
\end{equation}
where $\ket{x}$ is the register recording the index during the search, $\oplus$ is modulo $2$ addition, and $\ket{q}$ is ancilla qubit. When $f(x)=1$, the ancilla qubit will flip; otherwise, no operation is performed. Therefore, in the searching process of Grover's algorithm, we can prepare $\ket{x}\ket{0}$, then apply the oracle $\mathcal{O}$ and determine whether $x$ is the solution to the search problem by checking if the ancilla qubit is $\ket{1}$.

Researchers typically manually design these oracles using discrete quantum gates (i.e., quantum gates without parameters)~\cite{bijwe2022implementing,rahman2022grover}. However, with the proposal of AI-empowered search methods, parameterized quantum circuits have also been used to search for implementations of these oracles more efficiently. 
Search algorithms are also used for state preparation and operator synthesis~\cite{ashhab2022numerical,ashhab2024quantum}, evolve quantum oracles~\cite{ding2006evolving}, as well as design quantum adders~\cite{li2017approximate, deibuk2015design}, which is further applied to quantum autoencoders~\cite{lamata2018quantum}. 

\subsubsection{Quantum Architecture Search without Unitary Matrix}
In practical applications, situations without a target unitary matrix are more common, especially in quantum machine learning problems e.g. Quantum Neural Networks (QNN) or Variational Quantum Eigensolvers (VQE). Theoretically, if there are enough linearly independent input-output pairs (quantum initial states and corresponding final states), the unitary matrix corresponding to the transformation circuit is mathematically solvable, making this situation equivalent to the case with a target unitary matrix. However, the search process becomes more challenging when there is a lack of linearly independent input-output pairs.

The VQE algorithm~\cite{peruzzo2014variational} is a notable application without a target unitary matrix, with extensive utility in quantum chemistry, combinatorial optimization, and quantum simulation. Its core functionality involves optimizing parameterized quantum circuits through classical optimizers to determine the minimum eigenvalue and corresponding eigenvector of a given Hamiltonian. VQE has garnered significant attention in the NISQ era due to its relatively shallow parameterized quantum circuits and adaptability to noisy environments on quantum processors.
When utilizing VQE to solve quantum chemistry problems as the ground state energy estimation problem, the process typically involves discretizing electron distributions into orbitals through second quantization~\cite{berazin2012method}, mapping these orbitals to qubits using algorithms e.g. the Jordan-Wigner transformation~\cite{jordan1993paulische}, and employing unitary coupled cluster theory~\cite{taube2006new} to generate single and double excitation ansatze for quantum state evolution. This procedure necessitates the derivation of distinct quantum circuits for each molecule or Hamiltonian, requiring both quantum chemistry expertise and specialized algorithms. Moreover, the resulting ansatze are often deeper~\cite{barkoutsos2018quantum} due to the preservation of specific quantum chemical symmetries, rendering them challenging to implement on current NISQ hardware.

Consequently, researchers have explored QAS methods to automatically design VQE ansatz, aiming to reduce design complexity and generate circuits that satisfy given constraints (e.g., circuit depth, two-qubit gate count, or physical qubit connectivity). The search process also utilizes Hamiltonian expectation as the loss function. Existing research has applied QAS methods to enhance unitary coupled cluster ansatze~\cite{grimsley2019adaptive,sapova2022variational} or directly search for VQE circuits without prior ansatz knowledge~\cite{ostaszewski2021reinforcement,wang2022quantumnas,wu2023quantumdarts}.
Similarly, in combinatorial optimization problems, e.g. the maximum cut problem solved using the Quantum Approximate Optimization Algorithm (QAOA)\cite{farhi2014quantum}, studies have proposed methods to optimize the original QAOA circuit\cite{majumdar2021depth,majumdar2021optimizing} or design related circuits that circumvent Ising model decomposition~\cite{duong2022quantum,zhang2022differentiable}.

QNNs represent another application domain that does not require a specific target unitary matrix. Unlike VQE, which evolves a quantum state according to a given Hamiltonian, QNNs are a type of supervised learning algorithm. Inspired by classical neural network architectures~\cite{lecun2015deep,gu2018recent}, they employ layered quantum gate structures to emulate their classical counterparts. QNNs are trained to fit a dataset by minimizing a loss function, which is typically defined based on prediction accuracy during training. For instance, quantum convolutional neural networks~\cite{cong2019quantum} and quantum recurrent neural networks~\cite{bausch2020recurrent} incorporate manually designed quantum layers, implemented using parameterized and entanglement gates.
Similar to VQE, QNNs use parameterized quantum circuits, and their parameters are updated by a classical optimizer based on the calculated loss function. Consequently, these parameterized quantum circuits can be directly designed using QAS algorithms with the same original loss function. It is hypothesized that QNNs may offer advantages e.g. faster training and higher prediction accuracy compared to their classical counterparts~\cite{huang2022quantum}. Recent studies~\cite{zhang2021neural,wang2022quantumnas,duong2022quantum} have utilized QAS algorithms to either design complete QNN circuits or optimize specific layers, aiming to achieve improved training outcomes and modify predefined fixed ansatz templates like the Hardware Efficient Ansatz (HEA).

Beyond these applications, there are other potential applications without target unitary matrices, e.g. the design of quantum error correction codes and quantum nonlinear units. Quantum error correction codes are crucial in realizing fault-tolerant quantum computing by encoding quantum information to resist environmental noise~\cite{cai2021bosonic}. The essence of quantum error correction codes involves encoding the original quantum state into a larger state through specific encoding methods to enhance information protection. Consequently, automated circuit synthesis algorithms can be introduced in the encoding and decoding processes of quantum error correction codes. Current research in this domain encompasses heuristic methods~\cite{rigby2021heuristics}, machine learning approaches~\cite{chen2019machine,cong2019quantum}, and reinforcement learning techniques~\cite{nautrup2019optimizing, zeng2022approximate}.
Similarly, the design of quantum nonlinear units is also a potential application area. Since quantum circuits contain only linear unitary transformations, it is difficult for quantum machine learning to introduce nonlinearity as conveniently as in classical machine learning methods~\cite{schuld2014quest}. Some recent studies have attempted to use linear oracles to approximate the results of nonlinear operations~\cite{rattew2023non,guo2024nonlinear}, thus achieving the effect of nonlinear layers to some extent. This makes the design of nonlinear units another potential application area for QAS algorithms.

\subsection{Quantum Architecture Search Methods}
We summarize the various QAS algorithms in Table~\ref{tab:circuit_generation}. The table organizes the characteristics of QAS algorithms, particularly focusing on the methods for updating structural parameters and whether the updates of structural parameters and rotation parameters are based on gradient descent. Additionally, we indicate whether super-circuits are required during the search process and whether noise experiments are included. The specific algorithm classifications are as follows:

\begin{table*}[tb!]
\centering
\caption{Review of quantum architecture search (quantum logic circuit synthesis) methods. }
    \vspace{-17pt}
\resizebox{\textwidth}{!}{
\begin{threeparttable}
    \renewcommand\arraystretch{1.4}
    \label{tab:circuit_generation}
    \centering
    \begin{tabular}{c|c|c|c|c|c|c|c|c}
    \hline
    Paper & Method & Application & Sourced training & Structure Updating Method & Rotation Parameter Updating  & Super-circuit\tnote{1} Free & End-to-End\tnote{2} & Circuit Noise\\
    \hline
    \cite{williams1999automated} & Genetic Algorithm & Quantum Teleportation &\XSolidBrush & Genetic Algorithm & \XSolidBrush & \Checkmark & --- &  No\\
    \cite{khatri2019quantum} & Simulated Annealing & QFT &\XSolidBrush & Simulated Annealing & \Checkmark & \XSolidBrush & \XSolidBrush &  IBM-Q, Rigetti\\
    \cite{ye2021quantum} & RL & Bell state &\XSolidBrush & Reward Function & \XSolidBrush & \Checkmark & ---  & Simulated\\
    \cite{kuo2021quantum} & RL & Bell and GHZ state &\XSolidBrush & Reward Function & \XSolidBrush & \Checkmark & --- &  Simulated\\
    \cite{ostaszewski2021reinforcement} & RL & State Preparation &\XSolidBrush & Reward Function & \Checkmark & \Checkmark & \XSolidBrush  & No\\
    \cite{patel2024curriculum} & RL & State Preparation &\XSolidBrush & Reward Function & \Checkmark & \Checkmark & \XSolidBrush &  IBM-Q\\
    \cite{zhang2021neural} & Machine Learning & Max-Cut, Classification & \Checkmark & Neural Network & \Checkmark & --- & ---  & No\\
    \cite{he2024training} & Machine Learning & State Preparation &\XSolidBrush & Sampling & \Checkmark & \XSolidBrush& \XSolidBrush  & No\\
    \cite{zhang2022differentiable} & Machine Learning & Max-Cut, QFT, Error Mitigation &\XSolidBrush & Monte-Carlo Sampling & \Checkmark & \XSolidBrush & \Checkmark  & Simulated\\
    \cite{du2022quantum} & Machine Learning & State Preparation, Classification &\XSolidBrush & Sampling & \Checkmark & \XSolidBrush & \XSolidBrush  & Simulated\\
    \cite{wang2022quantumnas} & Machine Learning & State Preparation  &\XSolidBrush & Sampling & \Checkmark & \XSolidBrush & \XSolidBrush  & IBM-Q\\
    \cite{wu2023quantumdarts} & Machine Learning & State Preparation, Max-Cut, Classification & \XSolidBrush & Gumbel-Softmax & \Checkmark & \Checkmark & \Checkmark &  Simulated\\
    \hline
    \end{tabular}
    \begin{tablenotes}
        \small
        \item[1] pre-fixed circuit structure using human knowledge. \quad\quad   $^{\rm 2}$ the updating of both circuit structure and parameters are done by gradient descent.
    \end{tablenotes}

\end{threeparttable}}
\end{table*}

\subsubsection{Heuristic Algorithms}
Those particularly Genetic Algorithms (GAs), were among the earliest methods applied to automate quantum logic circuit synthesis~\cite{williams1999automated}. Initial research focused on using GAs to search for simple circuits, e.g. quantum teleportation. Subsequent studies demonstrated the potential of GAs in evolving relatively simple quantum circuits, including quantum error correction codes and quantum adders~\cite{massey2004evolving,bang2014genetic,las2016genetic,lamata2018quantum,potovcek2018multi}. The basic process of GAs involves: 1) selecting the initial population; 2) evaluating individuals in the current population; 3) sampling based on individual quality to form a candidate pool; and 4) performing crossover and mutation in the candidate pool. Steps 2 to 4 are repeated until a stop condition is met.

GAs are particularly suitable for problems where the solution is a sequence, as sequences can be viewed as strands of genes. From the quantum circuit representation methods introduced in the previous section, we can see that quantum circuits can be represented as sequences, with each gene locus corresponding to a quantum gate. Information e.g. gate type, parameters, and qubit index can be encoded as integer strings. The evaluation method in step 2 can refer to equations~\ref{eq:unitary} and~\ref{eq:non-unitary}, while crossover and mutation correspond to modifying the position, type, parameters, and qubit information of quantum gates. The performance of GAs is influenced by factors e.g. initial population size, candidate pool size (number of sampled quantum circuits), and gene length (quantum circuit depth). Increasing the types of gates in the circuit further extends the encoding length for each quantum gate, adding complexity to the optimization process.

In addition to genetic algorithms, simulated annealing algorithms have also been applied to QAS~\cite{khatri2019quantum}. This approach introduces parameterized quantum gates and randomly modifies parts of the existing optimal circuit structure during the search process. The internal parameters are then trained to obtain the optimal solution for the current structure. If this solution surpasses the existing optimal structure, an update is granted; otherwise, an update is made only with a probability that decreases exponentially with the performance gap.
While simulated annealing algorithms show some effectiveness, they suffer from weak scalability. Expanding the scale requires pre-defining a circuit structure, with each modification necessitating training to optimality, resulting in significant computational overhead. Consequently, heuristic algorithms have not found widespread application in the automatic synthesis and evolution of variational quantum circuits.

\subsubsection{Reinforcement Learning (RL) Algorithms}
To enhance the performance of QAS, researchers have explored machine learning algorithms beyond heuristic approaches, leveraging their powerful capabilities. RL, as a representative example of unsupervised learning, has emerged as a suitable alternative, given the difficulty of introducing labeled input-output pairs in QAS.
RL algorithms typically operate in discrete time steps, where an agent makes decisions at each step, and the environment provides corresponding rewards. The agent's goal is to maximize rewards through a limited set of actions. In QAS, the agent's action pool comprises candidate quantum gates, with each decision adding a gate to the circuit. The reward is calculated using a loss function after each addition. A key advantage of RL is its ability to generate circuits without preset maximum depths, theoretically allowing for continuous circuit deepening. Unlike heuristic methods, RL accommodates parameterized quantum gates, optimizing internal parameters while adding gates.

Primitive studies~\cite{kuo2021quantum,ye2021quantum} tested RL on two-qubit Bell states and three-qubit GHZ states. RL algorithms are then applied to automate the synthesis of variational quantum circuits for solving ground state energies~\cite{ostaszewski2021reinforcement}. Their experiments with four-qubit and six-qubit lithium hydride Hamiltonians assigned a -1 reward for each added gate and a large positive reward for achieving chemical accuracy (i.e., $1.6\times 10^{-3}$ Ha). This reward mechanism successfully guided the agent to find circuits achieving chemical accuracy, validating RL's feasibility.  \cite{patel2024curriculum} introduced curriculum learning, enhancing the guidance mechanism towards chemical accuracy and demonstrating significant performance improvements over previous work.
Following a similar RL-based paradigm, another line of research focused on applying these methods to variational quantum state diagonalization (VQSD). \cite{Diag1} proposed an RL-based approach to search for the optimal VQSD ansatz. While fundamentally similar to VQE, VQSD distinguishes itself by using the variance of the Hamiltonian as a loss instead of the energy expectation value. Built on this work,~\cite{Diag2} introduced a mechanism based on quantum information to accelerate the guidance of the RL agent, achieving a speedup of at least two-fold. Further advancements were made in~\cite{Diag3}, where the authors replaced the multi-layer perceptron in RL with a Kolmogorov-Arnold Network, demonstrating enhanced performance in the search for effective ansatze.

\subsubsection{Sampling-based Learning Algorithms}
In addition to reinforcement learning, sampling-based learning algorithms represent another significant class of circuit synthesis methods. These algorithms initially define the search space, considering a quantum circuit with $n$ qubits, $m$ layers, and $k$ candidate gates, resulting in a search space of size $k^{m\times n}$. The methodology proceeds as follows: first, sampling is performed on the $n\times m$ grid points to generate a complete quantum circuit. Subsequently, the sampled circuit undergoes evaluation, informing modifications to the sampling strategy for future iterations. This process is repeated iteratively to complete circuit synthesis. Notably, sampling-based learning algorithms demonstrate compatibility with parameterized quantum gates, allowing for optimization of internal circuit parameters post-sampling. The primary challenge inherent in this approach stems from the discrete nature of the sampling process, which precludes the differentiable gradients. Consequently, developing effective mechanisms for updating the sampling strategy remains a critical area of research in this domain.

\cite{zhang2022differentiable} first employed Monte Carlo gradient estimation methods to update sampling strategies, demonstrating efficacy in designing Quantum Fourier Transform ansatz and mitigating circuit errors. \cite{du2022quantum} selected the optimal circuit through multiple sampling and ranking, testing their algorithms on four-qubit hydrogen ground state energy prediction and image classification QNN design problems. \cite{wang2022quantumnas} proposed a method similar to~\cite{du2022quantum}, focusing on noise adaptation for quantum hardware. This approach generates shallow quantum circuits under the constraints of the quantum hardware to mitigate errors. Experimental results on IBM quantum devices demonstrated significant improvements over manually designed deeper circuits. However, both~\cite{du2022quantum} and~\cite{wang2022quantumnas} utilized pre-defined circuit modules (termed super-circuits or supernets), substantially reducing the search space through pre-designed modules or specified layer arrangements. This approach, however, may limit the algorithms' true capability, as their success heavily depends on these pre-designed circuit modules.

\cite{lu2023qas} addressed this issue by questioning the efficacy of circuit structure sampling and synthesis and aligning algorithm applications. The study proposed using the arbitrary unitary approximation problem to benchmark QAS algorithms, considering this task the most challenging in QAS. To accommodate potentially deep decomposition depths, a simplified version - arbitrary circuit reconstruction - was introduced. This task provides a unitary matrix with a known circuit solution within a certain depth, enabling fairer algorithm comparisons.
The results demonstrated that when prior information, e.g., pre-defined circuit modules, was removed, the performance of algorithms like~\cite{du2022quantum} and~\cite{zhang2022differentiable} decreased significantly. This highlights their strong dependence on well-chosen modules. Notably, when candidate gates were limited to non-parameterized quantum gates, machine learning algorithms generally underperformed compared to heuristic algorithms.

To enhance the efficiency and accuracy of searching algorithms,~\cite{wu2023quantumdarts} proposed a method utilizing Gumbel-Softmax~\cite{Gumbel1954,Bengio-gumbel-softmax16,gumbel-softmax16} to bridge the gap between discrete sampling and continuous gradient updates. This method uses the argmax term in the forward pass and the softmax term in backpropagation, addressing issues inherent in sampling-based learning algorithms and significantly improving search efficiency. The study eliminates prior knowledge interference by searching directly within the complete $k^{m\times n}$ space. It evaluates the method on three problems: VQE circuits (ground state energy prediction and Max-Cut) and QNN circuits (image classification), demonstrating notable improvements.
Furthermore, the article examines the scalability of QAS algorithms, proposing two search processes: Macro Search and Micro Search. Macro Search conducts global domain searches, while Micro Search uses preset sub-circuit modules with shared structural parameters but distinct internal rotation parameters. Unlike~\cite{du2022quantum} and~\cite{wang2022quantumnas}, which predetermine circuit module structures, \cite{wu2023quantumdarts} searches for the module structures. They facilitate the identification of frequently reused modules, e.g., single and double excitation operators in ground state energy estimation or quantum convolution and pooling modules in image classification. This approach reduces the search space and enables the application of modules discovered in small-scale problems (e.g., six-qubit lithium hydride) to larger-scale problems (e.g., 18-qubit methane). 

\subsubsection{Beyond Sampling-based}
Given that the aforementioned papers~\cite{wu2023quantumdarts,zhang2022differentiable} render the entire search process differentiable end-to-end, we can also categorize these types of methods as differentiable quantum architecture search. Inspired by the rapid development of generative models, particularly denoising diffusion models~\cite{sohl2015deep,rombach2022high}, one recent study~\cite{furrutter2024quantum} introduced a novel paradigm for quantum circuit synthesis. This approach bypasses the need for costly classical simulations of quantum dynamics by transforming the circuit synthesis problem into a generative one. Specifically, the authors encode quantum circuits into a continuous, high-dimensional tensor representation, allowing them to leverage powerful diffusion models to generate new circuits from noise. This method demonstrates an efficient and scalable way to explore the vast space of possible circuit configurations, offering a promising new direction for automated circuit design.
~\cite{zhang2021neural} proposed a neural network predictor for circuit synthesis, training it on small-scale quantum circuits and applying it to slightly larger-scale problems. This approach, while one of the few supervised learning models for QAS, faces limitations due to the diversity of quantum gates and the non-trivial relationship between gate configurations and circuit purposes. Recently, \cite{he2024training} introduced a training-free scheme based on~\cite{zhang2021neural}, which reverted to sampling-based optimization. However, this study's evaluation scale was limited (up to six qubits) and lacked comparison with state-of-the-art algorithms.

\section{Quantum Logic Circuit Optimization}\label{optimization}
Important factors need to be considered when executing quantum circuits, including qubit coherence time, environmental noise, etc. While quantum logic circuit synthesis generates circuits from scratch, optimization algorithms refine existing circuits based on specific targets. Consequently, circuits produced by synthesis algorithms can be further enhanced via optimization techniques. Moreover, by integrating the constraints of circuit synthesis with the optimization objectives, we can develop a unified synthesis optimization algorithm. It enables the direct creation of optimal quantum logic circuits that meet current optimization objectives.

\subsection{Problem Definition}
We begin by presenting a formulation of the quantum logic circuit optimization problem. Analogous to the quantum logic circuit synthesis problem, we employ a unitary matrix to represent a given quantum circuit. Let $\mathbf{U}$ denote the unitary matrix of the original circuit to be optimized, and $\hat{\mathbf{U}}$ represent the unitary matrix of the optimized circuit. Given a set of optimization targets $\kappa_i$, where we assume that all targets are to be minimized, the corresponding objective is:
\begin{equation}
    \min \sum_i \kappa_i(\hat{\mathbf{U}})  \quad \text{s.t.} \  ||\mathbf{U}-\hat{\mathbf{U}}||_F \leq \varepsilon,
\end{equation}
where $\varepsilon$ denotes the permissible unitary matrix error. When $\varepsilon$ equals $0$,  the two circuits are required to be completely equivalent. The term $\kappa_i(\hat{\mathbf{U}})$ represents the score of the optimized circuit with respect to a specific optimization target. In the following sections, we will summarize and categorize the primary optimization objectives of quantum logic circuits, followed by a systematic review of various optimization methodologies.

\subsection{Quantum Logic Circuit Optimization Targets}
The objectives of quantum logic circuit optimization algorithms are directly correlated to their optimization methods. Table~\ref{tab:circuit_optimization} illustrates the optimization methods in various research and their corresponding optimization objectives. It can be seen that, excluding brute-force search and reinforcement learning, other methods are aimed at more specific optimization targets. Overall, there is an emphasis on optimizing gate count and CNOT gates, while optimization of T gates is closely related to phase, resulting in a comparatively limited array of available optimization methods.
\begin{table*}
\caption{Quantum logical circuit optimization methods versus optimization target.}
   \vspace{-7pt}
\renewcommand \arraystretch{1.2}
\label{tab:circuit_optimization}
    \centering
    \begin{tabular}{|c|c|c|c|c|c|}
    \hline
    \multicolumn{2}{|c|}{\diagbox{Methods}{Target}} & Gate Count & Circuit Depth & CNOT Count & T Count / T Depth \\
    \hline
    \multicolumn{2}{|c|}{Brute-Force Search} &~\cite{kliuchnikov2013optimization,xu2022quartz}&~\cite{amy2013meet,kliuchnikov2013optimization,pointing2021optimizing}&\cite{bravyi20226}&\cite{amy2013meet}\\
    \hline
    \multirow{6}*{Pattern Matching}&Peephole Optimization &~\cite{maslov2016advantages,he2017decompositions,bravyi2021clifford} & & ~\cite{liu2021relaxed,sivarajah2020t,xu2023synthesizing}&~\cite{nam2020approximate} \\
    \cline{2-6}
    &Recursion & ~\cite{miller2011elementary,bae2020quantum}& & & \\
    \cline{2-6}
    &Divide and Conquer & ~\cite{prasad2006data}& & ~\cite{wu2020qgo,patel2021robust}& \\
    \cline{2-6}
    &Commute & ~\cite{hietala2021verified,bravyi2021clifford}& & ~\cite{nam2018automated}&~\cite{hietala2021verified,abdessaied2014quantum,zhang2019optimizing} \\
    \cline{2-6}
    &Template  &~\cite{bravyi2021clifford,maslov2003simplification,maslov2005quantum,maslov2008quantum,iten2022exact} & & & \\
    \cline{2-6}
    &Graph Contraction & & &~\cite{staudacher2023reducing} & ~\cite{amy2019t,staudacher2023reducing,kissinger2020reducing}\\
    \hline
    \multicolumn{2}{|c|}{Reduction}&~\cite{schneider2023sat}&&~\cite{schneider2023sat}&~\cite{heyfron2018efficient,amy2019t,amy2014polynomial}\\
    \hline
    \multirow{2}*{Machine Learning}&Heuristic Algorithm&&&~\cite{davis2019heuristics,amy2018controlled}&\\
    \cline{2-6}
    &RL&~\cite{fosel2021quantum,li2023quarl,riu2023reinforcement}&~\cite{fosel2021quantum,li2023quarl}&~\cite{li2023quarl,riu2023reinforcement}&~\cite{ruiz2025quantum}\\
    \hline
    \end{tabular}
\end{table*}
\subsubsection{Gate Count}\label{sec:optim_gatenum}
Optimization of the number of gates represents a fundamental objective in quantum circuit design. This optimization remains crucial across the current NISQ era and prospective FTQC paradigm. In the NISQ context, where qubit coherence time is limited and noise effects are significant, minimizing the number of quantum gates serves to reduce overall circuit noise, thereby enhancing operational accuracy. Moreover, in the realm of FTQC, gate count reduction contributes to improved computational efficiency of quantum hardware. Thus, the imperative to minimize quantum gates persists as a key strategy for advancing quantum circuit performance, irrespective of the underlying quantum computing framework. 
\cite{kliuchnikov2013optimization,prasad2006data,schneider2023sat} guaranteed optimal Clifford circuits, with~\cite{schneider2023sat} demonstrating scalability. Works are targeting other gate sets, e.g., the parameterized gate set~\cite{xu2022quartz}, Toffoli gates~\cite{maslov2003simplification}, Multi-Control Toffoli (MCT) circuits~\cite{miller2011elementary,bae2020quantum}, NCV circuits~\cite{maslov2005quantum,maslov2008quantum}, and parameterized VQE circuits~\cite{matsuzawa2020jastrow}.



\subsubsection{Circuit Depth}
It is intrinsically associated with the total runtime of a quantum circuit, which is constrained by the coherence time of qubits. Consequently, optimizing circuit depth holds practical significance for scalable quantum computing in the NISQ era. While reducing gate count and minimizing circuit depth are often considered alike, many studies address both objectives concurrently~\cite{kliuchnikov2013optimization,fosel2021quantum,li2023quarl}. However, these goals are not entirely identical. Depth optimization aims to enhance circuit parallelism, thereby reducing overall execution time and improving accuracy and efficiency. 

Circuit depth optimization is typically achieved by increasing the parallelism of quantum gates. Theoretically, a set of quantum gates without dependencies on each other can be executed in parallel. Dependencies in this context commonly refer to overlapping qubit operations. Thus, depth optimization can be understood as partitioning all gates in a circuit into the minimum number of mutually independent gate groups. 
Specifically, ~\cite{kliuchnikov2013optimization} achieved optimal circuits through recursion, a method that can be used to construct optimal circuit datasets.
~\cite{kattemolle2025edge} proposed to transform the quantum circuit depth optimization problem into solving the edge coloring problem for lattice graphs.
The algorithm presented in~\cite{amy2013meet} extended beyond depth optimization with the capability of various goals. 
With the use of redundant qubits,~\cite{sun2023asymptotically,yuan2023optimal} have yielded bounds for the optimal circuit depth, serving as a critical benchmark for our optimization pursuits. \cite{yuan2023does,yuan2025full} further explored how these bounds change under different physical qubit connectivities, which is crucial for tailoring optimization strategies to specific hardware architectures.
Other works focused on reducing the circuit depth of variational quantum circuits, e.g., QAOA circuits~\cite{herrman2021globally,herrman2022multi,chandarana2022digitized} and VQE circuits~\cite{kivlichan2018quantum}.

\subsubsection{CNOT Count}\label{sec:optim_cnot}
In addition to general gate optimization, targeting specific quantum gate types for reduction can significantly enhance circuit performance under certain conditions. In the NISQ era, the implementation of two-qubit entangling gates, e.g. CNOT gates, presents greater challenges compared to single-qubit gates. This is due to longer execution times (single-qubit gate: 60 ns, CNOT gate: 660 ns) and higher error rates (single-qubit gate $\approx 1\times 10^{-4}$, CNOT gate $\approx 1\times 10^{-2}$)\footnote{Data source: IBM-Q superconducting quantum computer Kyoto}. Consequently, minimizing the number of CNOT gates in circuit designs has emerged as a primary optimization objective, given that CNOT gates serve as representative two-qubit operations. This focused approach to gate reduction addresses the specific constraints and error profiles characteristic of current quantum hardware implementations. 

Numerous studies~\cite{barenco1995elementary,cybenko2001reducing,knill1995approximation,vartiainen2004efficient,mottonen2004quantum,shende2005synthesis,mottonen12006decompositions,shende2008cnot,bullock2008asymptotically} analyzed and optimized the theoretical number of CNOT gates in circuits by refining decomposition methods. These efforts have progressively reduced the theoretical upper bounds from \(\mathcal{O}(n^3 4^n)\) in~\cite{barenco1995elementary}, to \(\frac{23}{48}4^n - \frac{3}{2}2^n + \frac{4}{3}\) in~\cite{mottonen12006decompositions}. Regarding CNOT gate depth,~\cite{jiang2020optimal} demonstrated that any $n$-qubit CNOT circuit can be optimized to \(\mathcal{O}(\max\{\log n, \frac{n^2}{(n+m)\log(n+m)}\})\) depth using $m$ auxiliary qubits. A comprehensive discussion of recent advancements related to the theoretical number of CNOT gates is presented in~\cite{huang2023near}.

Recent studies~\cite{staudacher2023reducing,bravyi20226,schneider2023sat,davis2019heuristics,wu2020qgo,patel2021robust,liu2021relaxed,li2023quarl,riu2023reinforcement,gheorghiu2022reducing,nash2020quantum,amy2018controlled} focused on optimizing CNOT count in specific quantum circuits. \cite{bravyi20226} identified optimal Clifford circuits with minimal CNOT gates for up to 6 qubits, while~\cite{staudacher2023reducing} introduced methods to reduce CNOT count without compromising T gate optimization. Furthermore, optimizations tailored to VQAs aiming to enhance the performance in the NISQ era with~\cite{zhang2023low,yao2021adaptive,motta2021low,tkachenko2021correlation,chivilikhin2020mog,halder2024machine,anastasiou2024tetris,tang2021qubit,grimsley2019adaptive,magoulas2023linear} focused on optimizing VQE circuits, whereas~\cite{zhu2022adaptive,majumdar2021depth,majumdar2021optimizing} concentrated on QAOA circuit optimization. 

\subsubsection{T Count and T Depth}\label{sec:optim_tgate}
Unlike optimization goals in the NISQ era, fault-tolerant quantum computing (FTQC) implementation costs are not solely limited to hardware considerations, but also depend significantly on the temporal and spatial overheads of error-correcting codes. In FTQC models, Clifford group elements often have relatively simple transversal implementations across many error-correcting codes~\cite{zhou2000methodology}, whereas non-Clifford elements (e.g. T gates) are more complex to implement. For instance, in Steane codes~\cite{aliferis2005quantum} and surface codes~\cite{fowler2009high}, T gates require magic state distillation to implement~\cite{bravyi2005universal}. The space-time overhead for surface codes is analyzed in~\cite{campbell2017roads}, where Clifford gates incur a space-time cost of $\mathcal{O}(c_T d^3)$ with a constant factor $c_T$ of unit order, while T gates require $\mathcal{O}(C_T d^3)$ with $C_T \approx 160\sim 310$. Here, spatial overhead refers to the number of physical qubits needed to encode a logical qubit, and temporal overhead denotes the duration of the encoding protocol~\cite{campbell2017roads}. The implementation cost of T gates significantly exceeds that of Clifford gates by orders of magnitude~\cite{ruiz2025quantum,campbell2017roads}. Consequently, optimizing the quantity and depth of non-Clifford gates (T gates) is of paramount importance in FTQC.

Similar to reducing the overall gate count, directly minimizing the number of T gates is an intuitive approach~\cite{zhang2019optimizing,nam2020approximate,kissinger2020reducing,staudacher2023reducing,heyfron2018efficient,ruiz2025quantum,amy2019t,abdessaied2014quantum}.~\cite{abdessaied2014quantum} attempted to consolidate and reduce T count by reducing the number of H gates. ~\cite{nam2020approximate} optimized T count for approximate Quantum Fourier Transform circuits. \cite{kissinger2020reducing,heyfron2018efficient} employed the ZX diagram to reduce T count by extracting and consolidating non-Clifford phases. In~\cite{amy2019t}, the T count optimization problem was transformed into a minimum distance decoding problem for Reed-Muller codes, guaranteeing optimality.

However, as quantum hardware has evolved to support a greater degree of parallel execution of gates, the optimization objective has shifted from minimizing T count to minimizing the number of sequential T-gate layers, known as T depth. This metric is more crucial for reducing the overall circuit execution time. Accordingly, studies have focused on exploiting T-gate parallelism to minimize T depth in quantum circuits~\cite{zhang2019optimizing,amy2013meet,amy2014polynomial}. These investigations primarily examined the effect of parallelizable T gates on reducing overall circuit operation time. Notably,~\cite{amy2014polynomial} proposed a polynomial-time optimization algorithm that, in most instances, guarantees the resultant circuit exhibits minimal T depth.

\subsection{Quantum Logic Circuit Optimization Methods}
Based on the characteristics of different optimization methods, we summarize them in Table~\ref{tab:circuit_optimization_feature}. The table lists the optimization targets and circuit representation methods for each optimization algorithm. It can be seen that gate circuits dominate, while phase polynomial representations are mainly used for T gate optimization. We also list some specific features, e.g., whether the algorithm requires ancillary qubits, whether it provides optimal solutions, and whether it optimizes parameterized circuits.
\begin{table*}
    \centering
    \caption{Review of quantum logical circuit optimization algorithms and characters.}
    \vspace{-7pt}
    \renewcommand\arraystretch{1.2}
    \resizebox{0.9\textwidth}{!}{
    \begin{tabular}{c|c|c|c|c|c|c|c|c|c}
    \hline
    \multirow{2}{*}{Reference} & \multirow{2}{*}{Target} & \multicolumn{5}{c|}{Circuit Representation} & \multirow{2}{*}{Ancilla qubits} & \multirow{2}{*}{Is Optimal} & \multirow{2}{*}{Parameterized Circuit} \\
    \cline{3-7}
    &&Gate Model&Phase Polynomial&DAG&ZX Diagram&Tensor Network&&&\\
    \hline
    \cite{prasad2006data}&Gate Count&\Checkmark&&\Checkmark&&&\XSolidBrush&\XSolidBrush&\XSolidBrush\\
    \hline
    \cite{xu2022quartz}&Gate Count&&&\Checkmark&&&\XSolidBrush&\XSolidBrush&\Checkmark\\
    \hline
    \cite{iten2022exact}&Gate Count&&&\Checkmark&&&\XSolidBrush&\XSolidBrush&\Checkmark\\
    \hline
    \cite{maslov2016advantages}&Gate Count&\Checkmark&&&&&\Checkmark&\XSolidBrush&\XSolidBrush\\
    \hline
    \cite{he2017decompositions}&Gate Count&\Checkmark&&&&&\Checkmark&\XSolidBrush&\XSolidBrush\\
    \hline
    \cite{miller2011elementary}&Gate Count&\Checkmark&&&&&\Checkmark&\XSolidBrush&\XSolidBrush\\
    \hline
    \cite{bae2020quantum}&Gate Count&\Checkmark&&&&&\XSolidBrush&\XSolidBrush&\XSolidBrush\\
    \hline
    \cite{bravyi2021clifford}&Gate Count&\Checkmark&&&&&\XSolidBrush&\XSolidBrush&\XSolidBrush\\
    \hline
    \cite{maslov2003simplification}&Gate Count&\Checkmark&&&&&\XSolidBrush&\XSolidBrush&\XSolidBrush\\
    \hline
    \cite{maslov2005quantum}&Gate Count&\Checkmark&&&&&\XSolidBrush&\XSolidBrush&\XSolidBrush\\
    \hline
    \cite{maslov2008quantum}&Gate Count&\Checkmark&&&&&\XSolidBrush&\XSolidBrush&\XSolidBrush\\
    \hline
    \cite{fosel2021quantum}&Gate Count, Circuit Depth&\Checkmark&&&&\Checkmark&\XSolidBrush&\XSolidBrush&\XSolidBrush\\
    \hline
    \cite{kliuchnikov2013optimization}&Gate Count, Circuit Depth&\Checkmark&&&&&\XSolidBrush&\Checkmark&\XSolidBrush\\
    \hline
    \cite{li2023quarl}&Gate Count, Circuit Depth, CNOT Count&&&\Checkmark&&&\XSolidBrush&\XSolidBrush&\XSolidBrush\\
    \hline
    \cite{riu2023reinforcement}&Gate Count, CNOT Count&&&&\Checkmark&&\XSolidBrush&\XSolidBrush&\XSolidBrush\\
    \hline
    \cite{schneider2023sat}&Gate Count, CNOT Count&\Checkmark&&&&&\XSolidBrush&\Checkmark&\XSolidBrush\\
    \hline
    \cite{hietala2021verified}&Gate Count, T Count/Depth&\Checkmark&&&&&\XSolidBrush&\XSolidBrush&\Checkmark\\
    \hline
    \cite{pointing2021optimizing}&Circuit Depth&\Checkmark&&&&&\XSolidBrush&\XSolidBrush&\XSolidBrush\\
    \hline
    \cite{amy2013meet}&Circuit Depth, T Count/Depth&\Checkmark&&&&&\Checkmark&\Checkmark&\XSolidBrush\\
    \hline
    \cite{nam2018automated}&CNOT Count&\Checkmark&\Checkmark&\Checkmark&&&\XSolidBrush&\XSolidBrush&\Checkmark\\
    \hline
    \cite{bravyi20226}&CNOT Count&\Checkmark&&&&&\XSolidBrush&\Checkmark&\XSolidBrush\\
    \hline
    \cite{davis2019heuristics}&CNOT Count&\Checkmark&&&&&\XSolidBrush&\XSolidBrush&\Checkmark\\
    \hline
    \cite{wu2020qgo}&CNOT Count&\Checkmark&&&&&\XSolidBrush&\XSolidBrush&\Checkmark\\
    \hline
    \cite{patel2021robust}&CNOT Count&\Checkmark&&&&&\XSolidBrush&\XSolidBrush&\Checkmark\\
    \hline
    \cite{xu2023synthesizing}&CNOT Count&\Checkmark&&&&&\XSolidBrush&\XSolidBrush&\Checkmark\\
    \hline
    \cite{amy2018controlled}&CNOT Count&&\Checkmark&&&&\Checkmark&\XSolidBrush&\XSolidBrush\\
    \hline
    \cite{sivarajah2020t}&CNOT Count&&&\Checkmark&&&\XSolidBrush&\XSolidBrush&\Checkmark\\
    \hline
    \cite{liu2021relaxed}&CNOT Count&\Checkmark&&&&&\Checkmark&\XSolidBrush&\XSolidBrush\\
    \hline
    \cite{staudacher2023reducing}&CNOT Count, T Count/Depth&&&&\Checkmark&&\XSolidBrush&\XSolidBrush&\XSolidBrush\\
    \hline
    \cite{abdessaied2014quantum}&T Count/Depth&&\Checkmark&&&&\XSolidBrush&\XSolidBrush&\XSolidBrush\\
    \hline
    \cite{zhang2019optimizing}&T Count/Depth&&&\Checkmark&&&\Checkmark&\XSolidBrush&\XSolidBrush\\
    \hline
    \cite{nam2020approximate}&T Count/Depth&\Checkmark&&&&&\Checkmark&\XSolidBrush&\Checkmark\\
    \hline
    \cite{kissinger2020reducing}&T Count/Depth&&\Checkmark&&\Checkmark&&\XSolidBrush&\XSolidBrush&\Checkmark\\
    \hline
    \cite{heyfron2018efficient}&T Count/Depth&&&&&\Checkmark&\Checkmark&\XSolidBrush&\XSolidBrush\\
    \hline
    \cite{ruiz2025quantum}&T Count/Depth&&&&&\Checkmark&\Checkmark&\Checkmark&\XSolidBrush\\
    \hline
    \cite{amy2014polynomial}&T Count/Depth&&\Checkmark&&&&\XSolidBrush&\XSolidBrush&\XSolidBrush\\
    \hline
    \cite{amy2019t}&T Count/Depth&&\Checkmark&&&&\Checkmark&\XSolidBrush&\XSolidBrush\\
    \hline
    \end{tabular}}
    \label{tab:circuit_optimization_feature}
\end{table*}


\subsubsection{Brute Force Search (BFS)}\label{bruteforce}
BFS finds optimal circuits with specific depth and width through exhaustive enumeration. While this method guarantees optimal results, it incurs significant computational costs, limiting its application to small-scale logic circuits. Nevertheless, the optimal circuits generated by BFS can be further utilized in other optimization algorithms. 
The BFS method typically comprises two core steps. First, a dataset containing all quantum logic circuits meeting specific criteria (e.g., certain depth and width) is constructed based on a set of candidate quantum gates. Subsequently, this dataset is traversed to identify the optimal circuit. As every subcircuit of an optimal quantum circuit is also optimal, the second step can employ bidirectional search or recursive backtracking to gradually identify the optimal quantum circuit composing the target unitary matrix from the dataset. However, these algorithms offer limited improvement in search efficiency. Therefore, investigating the optimization of dataset construction to reduce dataset volume and consequently decrease computational costs associated with searching is equally crucial.

The work~\cite{amy2013meet} adopted a breadth-first search scheme to construct a gate circuit tree, where each edge denotes the addition of a gate. The study used a bidirectional search scheme to reduce the search depth, enabling the discovery of larger circuits from smaller datasets (e.g., finding optimal circuits up to depth $2k$ using a dataset with depth $k$). The bidirectional approach relies on the principle that subcircuits of optimal circuits are also optimal.
Another contemporary study~\cite{kliuchnikov2013optimization} proposed using unitary matrix equivalence classes to reduce the dataset size. By grouping unitary matrices that are equivalent under logical qubit reordering, they significantly compressed the search space. Building on this idea,~\cite{bravyi20226} investigated the construction of equivalence classes based on CNOT depth, considering nine different scenarios for adding CNOT gates (accounting for various placements of single-qubit H and P gates). It groups unitary matrices that are equivalent after removing single-qubit gates and reordering qubits, allowing the search for all optimal circuits within six qubits by first reducing a target unitary to its representative element.

The optimization methods employed in~\cite{xu2022quartz} and~\cite{pointing2021optimizing} were analogous to breadth-first search. They first constructed a dataset of optimal subcircuits and then reduced circuit costs by replacing local segments of the target circuit with optimal ones. However, this approach only guaranteed local optimality, not global optimality. The algorithm proposed in~\cite{pointing2021optimizing} initially defined an $n$-qubit, $m$-layer grid with randomly placed quantum gates, from which equivalence classes were then constructed. For a target circuit, all replaceable subcircuits were identified, and optimal replacements with minimal depth were sought from the dataset's equivalence classes. Extending this approach,~\cite{xu2022quartz} applied it to continuous quantum gate sets. The research also introduced a verification method for parameterized circuits using the function $|\bra{\psi_0}\mathbf{U}(\mathbf{p}_0)\ket{\psi_1}|$, where $\mathbf{U}(\mathbf{p}_0)$ represents a quantum circuit with parameters $\mathbf{p}_0$, $\ket{\psi_0}$ and $\ket{\psi_1}$ are random quantum states. By randomly sampling the states and parameters, two unitary matrices were recognized as equivalent if they yielded identical function values, enabling the optimal replacement of local subcircuits within parameterized circuits.

\subsubsection{Pattern Matching}
It is a popular technique in quantum circuit optimization that identifies and replaces known patterns within a circuit for efficiency. While it typically focuses on optimizing local sub-circuits and does not guarantee global circuit optimality, it is favored for its low time complexity and scalability.
The process of pattern matching begins by establishing a set of sub-circuit substitution rules. These rules are applied by traversing sub-circuits, often using methods like circuit partitioning or sliding windows. When applicable, substitution rules are executed to minimize sub-circuit cost.

\textbf{Peephole Optimization:}
Peephole optimization, a classical compilation technique~\cite{mckeeman1965peephole}, was applied to small segments, e.g. a few instructions in classical computing or a limited number of quantum gates. In quantum circuit optimization, it identified local patterns that could be optimized by traversing the circuit.
A specialized form based on qubit states~\cite{liu2021relaxed} analyzed qubit states using quantum state annotation, akin to finite state machines in classical compilers. Optimization was applied when a qubit was found in a predefined pure or basis state, and a local sub-circuit matched a predefined pattern. This method altered the unitary matrix of the optimized circuit while maintaining functionality. For example,~\cite{maslov2016advantages,he2017decompositions} optimized multi-control NOT gates by replacing Toffoli gates with phase Toffoli gates to reduce implementation costs. These phase Toffoli gates differed from the standard ones only by a phase factor on certain states, and the global effect was equivalent to that of a Toffoli gate, with reduced cost.
The $\rm t\ket{ket}$ compiler incorporated peephole optimization as a subroutine~\cite{sivarajah2020t}, targeting specific quantum patterns. Symbolic peephole optimization~\cite{bravyi2021clifford} improved upon traditional pattern matching, addressing limitations where replacement sub-circuits must be fully decoupled from the rest of the circuit. This approach decomposed CNOT gates into projection operators on control qubits and symbolic Pauli operators on target qubits, optimizing with dynamic programming and predefined equivalence rules.

\textbf{Recursive Decomposition:}
Recursive methods were often used to identify patterns in quantum circuits, particularly for multi-controlled NOT gates. These methods broke down a controlled-NOT gate circuit with 
$c$ control qubits into circuits with $c_1$ and $c_2$ control qubits, where $c=c_1+c_2$. A common decomposition approach sets $c_1=c-1$ and $c_2=1$, reducing one control qubit per iteration. In~\cite{bae2020quantum}, this method replaced a controlled-NOT gate with $c$ control qubits by one with $c-1$ qubits and additional CNOT gates, verified using quantum Karnaugh maps. The approach in~\cite{miller2011elementary} generalized this strategy, considering all possible pairings of $(c_1,c_2)$, generating multiple combinations and selecting the most efficient one.

\textbf{Divide-and-Conquer:}
it splits circuits into sub-circuits of specified depth and width, which are optimized independently before being recombined. The QGo algorithm~\cite{wu2020qgo} optimized CNOT and SWAP gates by partitioning an 
$n$-qubit circuit into multiple $k$-qubit blocks. It employed a greedy heuristic based on a quantum gate dependency graph, selecting the best partitions and iterating until the entire circuit was decomposed. QGo then used optimal circuit synthesis~\cite{davis2020towards} to reconstruct the sub-circuits. A similar approach in~\cite{prasad2006data} employed a linear, array-like data structure to describe quantum circuits, which can be viewed as the result of a depth-first search traversal of the DAG. Building upon~\cite{wu2020qgo}, 
QEst~\cite{patel2021robust} replaced exact synthesis with approximate synthesis, achieving simpler circuit implementations with acceptable error margins.

\textbf{Commutation:}
Commutation-based methods optimized circuits by rearranging commutable gates. The works in~\cite{abdessaied2014quantum} and~\cite{nam2018automated} focused on optimizing the Clifford+T set by reducing the number of T gates through phase polynomials. H gates, which interfered with phase merging, were rearranged to cancel out and minimize their count. \cite{bravyi2021clifford} utilized commutation rules to optimize Pauli and SWAP gates by partitioning the circuit into three sections: Pauli gates, SWAP gates, and computational components (including CNOT, H, and S gates), optimizing each with tailored strategies. Additionally,~\cite{hietala2021verified} introduced SQIR (Small Quantum Intermediate Representation) for efficient quantum program verification and equivalence checking, combining commutation with gate elimination.

\textbf{Template Matching:} \label{template}
it extends pattern matching by optimizing sequences of gates that result in the identity transformation, where the first half of the sequence is the inverse of the second. For a template consists of $m$ gates that satisfy $\prod_{i=1}^{m}\mathbf{U}_i = \mathbf{I}$, if a segment in the circuit matches the first $k$ gates of the template, it can be replaced by the inverse of the remaining gates $\mathbf{U}_1\mathbf{U}_2\cdots\mathbf{U}_k \rightarrow \mathbf{U}_m^{-1}\mathbf{U}_{m-1}^{-1}\cdots\mathbf{U}_{k+1}^{-1}$. The optimization reduces circuit length as long as $k \ge \left\lfloor \frac{m}{2} \right\rfloor + 1$.

\cite{maslov2003simplification} employed template matching to optimize circuits containing NOT, CNOT, Toffoli, and MCT gates. They explored all templates of length $\leq 7$ and proposed strategies for optimization by traversing circuits bidirectionally to identify matching segments. Similar methodologies were applied in~\cite{maslov2005quantum} with optimizations tailored to specific gate sets, and explored additional circuit rules, e.g. gate commutativity.
Template matching algorithms can also optimize gate count and circuit depth~\cite{maslov2008quantum}, leveraging gate commutativity to enhance parallelism of gates. \cite{bravyi2021clifford}, meanwhile, employed commutation and template matching to optimize CNOT, H, and S gates.
\cite{bravyi2021clifford} introduced DAG-based template matching with templates as subgraphs. A bidirectional search finds the match for both predecessor and successor nodes in the DAG.

\textbf{Graph Contraction:}
Primarily optimized circuits represented by ZX diagrams, targeting specific local patterns. Several representative patterns include:
1) \textit{Local Complementation}: Merging elements with special angles (e.g., \( \pm \frac{\pi}{2} \)) into adjacent angles.
2) \textit{Pivoting}: Combining multiples of 0 or \( \pi \) through rotation into adjacent nodes.
3) \textit{Elimination and Fusion}: Elimination of Empty Units and Fusion of Adjacent Units in Clifford Circuits. 

In~\cite{kissinger2020reducing}, phase optimization in ZX diagrams reduced T count after circuit decomposition. For non-Clifford phases, phase gadgets were introduced and isolated using H gates. Rule-based contraction optimization are then conducted to reduce the T count.
Similarly,~\cite{staudacher2023reducing} minimized CNOT count without compromising T count by using a heuristic scoring system, allowing random or greedy selection of optimization strategies. The scoring system can be adapted to different objectives by modifying the criteria. The study emphasizes that indiscriminate fusion of adjacent units may not always yield positive outcomes; conversely, dismantling units might facilitate the execution of new rules.

\subsubsection{Reduction}
An alternative approach to addressing optimization problems is to reduce them to other problems and apply specialized solvers. A common method utilizes phase polynomials to represent quantum circuits, which are then converted into Reed-Muller codes. This transformation reduces the circuit optimization problem to a Reed-Muller decoding problem, enabling the use of dedicated solving techniques.

Phase polynomials are often employed in optimizing T gates. In a phase function, as shown in Eq.~\ref{eq:phasepoly}, introducing a common factor of $\frac{\pi}{4}$ ensures that each coefficient $\theta_i \in \mathbb{Z}_8$, where the coefficients represent the rotation parameters of Rz gates in the range $[0, 2\pi]$. Even coefficients indicate that the current phase can be realized using S gates, meaning the number of T gates directly correlates with the number of odd coefficients. When these coefficients are further reduced modulo 2, where 1 and 0 represent odd and even coefficients, respectively, we seek a set $\delta\bm\theta\in\{0,1\}^{2^n-1}$ such that $\bm{\theta} \oplus \delta \bm{\theta}$ is equivalent to $\bm{\theta}$ in the circuit (where $\oplus$ denotes bitwise XOR), with the Hamming weight of the new circuit minimized.
This problem is equivalent to finding a binary code in an equivalence class that has the smallest Hamming distance from $\bm{\theta}$. In~\cite{amy2019t}, it is shown that for circuits containing only T and CNOT gates, the required code size is $\mathcal{RM}(n-4, n)^*$, providing comprehensive theoretical proofs for related complexity bounds. \cite{heyfron2018efficient} proposed a compilation optimization paradigm for Clifford+T sets, specifically targeting T gate optimization. Non-Clifford phases were first extracted by introducing redundant qubits, isolating the Clifford parts in the circuit. The optimization of non-Clifford phases was then reduced to Reed-Muller (RM) decoding and low-rank decomposition problems. Prior to reducing to RM code decoding, \cite{amy2014polynomial} also attempted to reduce the optimization problem for T gates to quasi-matrix partitioning problems, using the same problem setup as~\cite{amy2019t} but constructing the phase function set as a finite quasi-matrix.

Another popular method is to reduce the circuit optimization problem to a SAT problem for resolution. \cite{schneider2023sat} utilized the conclusion from~\cite{aaronson2004improved} that an optimal Clifford circuit requires at most $\mathcal{O}(\frac{n^2}{\log n})$ quantum gates, focusing on the synthesis and optimization of Clifford circuits. Assuming the optimal circuit contains $m$ gates, where $0 < m \leq c\frac{n^2}{\log n}$ and $c$ is a constant, at each step $k \leq m$, there are $\mathcal{O}(n^2)$ ways to add quantum gates to an existing circuit. This implies a total of $m(1 + n + n^2)$, i.e., $\mathcal{O}(\frac{n^4}{\log n})$ binary variables are needed to represent the composition of an optimal Clifford circuit, thereby transforming the circuit optimization problem into a SAT problem for resolution. To ensure the optimal $m$, a binary search strategy can be employed to iterate through the possible values of $m$, given that $m$ ranges from $1$ to $\mathcal{O}(\frac{n^2}{\log n})$.  

\subsubsection{Machine Learning}\label{ML}

In complement to traditional algorithms, machine learning and associated techniques have been incorporated into quantum logic circuit optimization. Although machine learning algorithms may not provide theoretical guarantees on solution accuracy, they demonstrate the capability in addressing larger-scale problems, offering feasible solutions at reduced computational costs. This attribute imparts practical value to quantum logic circuit optimization algorithms, expanding their applicability in more complex scenarios.

\textbf{Heuristic Algorithms:}
The principle of heuristic algorithms is to search the solution space through a series of rules and heuristic strategies to find near-optimal solutions. Compared to traditional exact algorithms, heuristic algorithms are generally more efficient, especially when dealing with quantum logic circuit optimization problems with large solution spaces, offering near-optimal solutions at lower computational costs. The work~\cite{amy2018controlled} proposed a relatively primitive heuristic algorithm to optimize CNOT counts in phase polynomial representations. It constructs a ${0,1}$ matrix based on XOR strings in the phase function (see Eq.~\ref{eq:phasepoly}), where each column represents an XOR string from the phase function, with matrix elements set to 1 for corresponding qubits in the XOR string.
Adding CNOT gates causes XOR flips in the control qubit row based on the target qubit row. At each step, the algorithm greedily selects the qubit pair that causes the most flips to add a CNOT gate. If a column has only two $1$, it deterministically adds that CNOT gate. This constructs a pure CNOT circuit and analyzes the optimal CNOT count solution. \cite{davis2019heuristics} employed the $A^{*}$ algorithm to optimize CNOT counts in circuits. Given a set of quantum gates, allowed precision error, and maximum CNOT count, the algorithm starts from an empty circuit. Candidate circuits are stored in a priority queue. In each iteration, the algorithm pops the top circuit, attempts to append gates from the given set to the circuit's end, calculates the distance to the target unitary, and checks if it's within the allowed error range or if the CNOT count exceeds the limit. If requirements are met, it returns the current circuit and terminates. If the queue is empty, no solution exists. The comparison function of the priority queue is the same as the total cost of $A^{*}$, comprising the cost of the existing path and a heuristic estimate of the additional cost to reach the target unitary.

\textbf{Reinforcement Learning:}
In quantum logic circuit optimization, reinforcement learning offers a significant advantage over other algorithms by automatically discovering novel or undefined optimization strategies and matchable patterns. \cite{fosel2021quantum} presented a preliminary algorithm that encodes quantum gates as triplets within a three-dimensional tensor representation of quantum circuits, using qubit indices, gate types, and depth as coordinates. The algorithm incorporates predefined hard rules (transformations with direct benefits, e.g., adjacent gate cancellation) and soft rules (transformations with indirect potential benefits, like swapping adjacent commutable gates). During each iteration, the agent is limited to deciding on soft rule usage, while hard rules are applied to the maximum extent for optimal benefit. Agent actions are tensor-encoded, with single-gate rules encoded by depth and qubit index and two-gate rules by the minimum common qubit index and the depth of the first gate. The reinforcement learning-trained agent employs a convolutional neural network, taking the circuit's 3D tensor representation as the input state and outputting the policy (encoded circuit transformation actions) and the current state value (indicating optimization potential). Rewards are based on cost changes, e.g. the reduction in the CNOT count, before and after each transformation.

The study in~\cite{li2023quarl} replaced manually specified action sets with a policy network. It first converts gate circuits into graph structures and employs a graph neural network to extract quantum gate representations. The policy network includes quantum gate and transformation selection, incorporating probability distribution-based sampling with gradient updates via temperature-coefficient softmax. Experimental results show significant improvements over existing compilation optimizers.
In~\cite{riu2023reinforcement}, a reinforcement learning method is introduced for ZX calculus, using a policy network in combination with a graph neural network to identify replaceable positions in ZX diagrams. 
Actions involve substituting the output positions from the policy network, where the substitution concentrates on common equivalent circuit simplifications for the Clifford set. Rewards are based on gate count, necessitating the conversion of ZX diagrams to quantum circuits. Delayed conversion strategies are employed to reduce computational overhead.
~\cite{ruiz2025quantum} optimized T count by transforming the problem into the low-rank decomposition of three-dimensional tensors, adopting Google's AlphaTensor method~\cite{fawzi2022discovering}. It employed phase polynomials up to the third order, corresponding to T gates, Controlled-S gates, and Controlled-Controlled-Z gates. By encoding these coefficients into a third-order tensor, the problem was reframed as the optimal decomposition of the circuit's third-order tensor. Additional pattern matching strategies for T and Toffoli gates were employed to streamline consecutive circuit segments. The study highlights that reinforcement learning algorithms can spontaneously discover novel matching patterns, complementing the manually designed strategies.

\subsubsection{Circuit Optimization for VQAs}
VQAs are considered among the most promising candidates for demonstrating quantum supremacy in the NISQ era, owing to their shallow circuit depth and the adaptability to circuit noise and other factors. Consequently, numerous quantum circuit optimization algorithms have been developed, targeting specific VQAs and corresponding quantum hardware characteristics. Several comprehensive reviews have explored related issues~\cite{tilly2022variational,fedorov2022vqe,cerezo2021variational,blekos2024review}. This section presents a concise overview.

\textbf{Variational Quantum Eigensolver:}
A key application of VQE is solving ground state energies in quantum chemistry problems. Common algorithms include HEA~\cite{kandala2017hardware} and Unitary Coupled Cluster (UCC)~\cite{romero1701strategies}. HEA circuits adhere to hardware constraints and are simple but less accurate, while UCC methods are precise but complex due to creation-annihilation operator implementations. Continuous efforts have been focused on simplifying these operators in circuits.
ADAPT-VQE~\cite{grimsley2019adaptive} reduced operator usage by incrementally building ansatz circuits, selecting operators with the highest energy gradient from a pre-determined pool. This accelerates VQE convergence and significantly reduces the number of required operators and circuit depth. Qubit-ADAPT-VQE~\cite{tang2021qubit} implemented a hardware-friendly ADAPT-VQE variant to minimize SWAP gates in existing quantum hardware topologies.
QEB-ADAPT-VQE~\cite{yordanov2021qubit} used Qubit-excitation evolution instead of Fermionic-excitation, allowing fixed structures for single and double excitation operators. It outperforms FEB-based algorithms~\cite{grimsley2019adaptive,tang2021qubit} in CNOT count, circuit parameter count, and algorithm iterations. \cite{magoulas2023linear} combined FEB and QEB methods, employing an approximation strategy to balance CNOT gate reduction and accuracy preservation. DISCO-VQE~\cite{burton2023exact} achieved more compact ansatz circuits by combining spin-symmetry-preserving fermionic operator products with a global optimization algorithm, simultaneously optimizing discrete operators and continuous variational parameters. TETRIS-ADAPT-VQE~\cite{anastasiou2024tetris} accelerated ansatz synthesis by introducing multiple non-interfering operators per iteration, contrasting with ADAPT-VQE's single-operator approach.
Diverging from ADAPT methods, \cite{halder2024machine} incorporated Restricted Boltzmann Machines (RBM)~\cite{hinton2012practical} and many-body perturbation theory in ansatz construction. It utilized low-level excitation UCC operators as the RBM training set to generate higher-level excitation operators, achieving high accuracy while maintaining low complexity (e.g., UCCSDT accuracy with UCCSD-level circuit complexity).

\textbf{Quantum Approximate Optimization Algorithm (QAOA):}
As introduced by~\cite{farhi2014quantum}, it plays a crucial role in solving quadratic unconstrained binary optimization (QUBO) problems. \cite{alam2020circuit} proposed four general methods for optimizing QAOA circuits through gate commutation. Focusing specifically on QAOA circuits designed to solve the Max-Cut problem,~\cite{arufe2022quantum} presented a decomposition-based genetic algorithm for circuit optimization.
The work~\cite{majumdar2021optimizing} introduced two optimization strategies based on edge coloring~\cite{west2001introduction,vizing1964estimate} and depth-first search, respectively, yielding optimizations of $\lfloor \frac{n}{2} \rfloor$ and $n-1$ CNOT gates (where $n$ represents the number of vertices in the original graph). \cite{majumdar2021depth} further improved upon the DFS-based strategy from~\cite{majumdar2021optimizing}, proposing a heuristic greedy algorithm that reduced circuit depth while maintaining the $n-1$ CNOT gate optimization.
Inspired by the ansatz construction in ADAPT-VQE~\cite{grimsley2019adaptive,tang2021qubit}, ADAPT-QAOA~\cite{zhu2022adaptive} selected operators with the highest energy gradient from an operator pool in each iteration, accelerating algorithm convergence. ADAPT-QAOA can be adapted to various hardware constraints, increasing the approximation ratio while reducing circuit parameters and CNOT count. Furthermore, Dynamic-ADAPT-QAOA~\cite{yanakiev2024dynamic} built upon ADAPT-QAOA by eliminating layers with optimal parameters approaching zero, further reducing QAOA circuit size.
In~\cite{herrman2021globally}, the authors approached QAOA circuit optimization from the perspective of enhancing combinatorial optimization problems, providing a global optimization method.
Another aspect in optimizing the QAOA ansatz is to improve the expressivity of the single QAOA layer and thus reduce the overall layer numbers to optimize circuit depth. MA-QAOA~\cite{herrman2022multi} disabled the parameter sharing strategy, leading to qubit-number independent parameters for each layer. XQAOA~\cite{XQAOA} altered the mixer Hamiltonian with extra parameterized rotation gates to further enhance ansatz expressivity.

\section{Qubit Mapping and Routing}
\begin{figure*}
    \centering
    \includegraphics[width=1\textwidth]{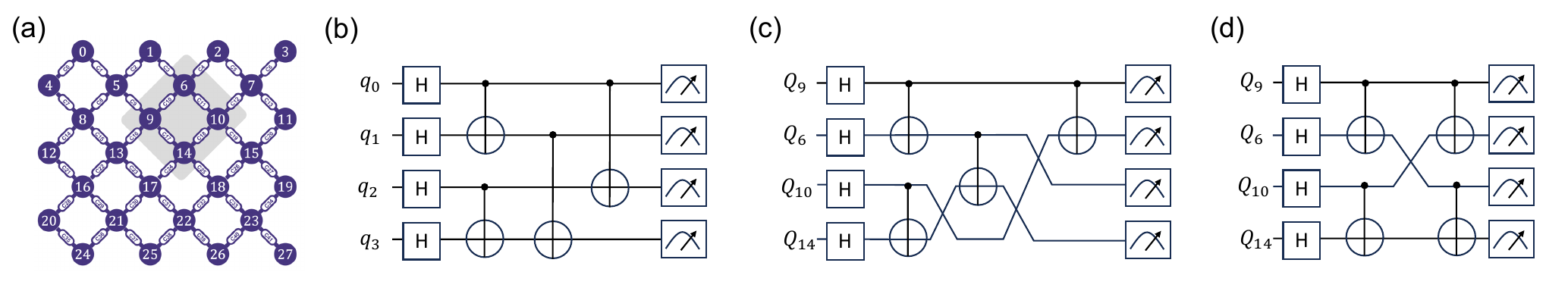}
    \vspace{-15pt}
    \caption{Quantum Circuit Compiling. (a) An example qubit topology of a superconducting quantum processor, we select four working qubits (marked as grey); (b) Logical circuit; (c) After mapping logical qubits $q_0,q_1,q_2,q_3$ to physical qubits $Q_9,Q_6,Q_{10},Q_{14}$, three SWAP gates are inserted since only neighbor qubits are connected; (d) Circuit after qubit routing which reduce the number of SWAP gates to only one.}
    \label{fig:compiling}
\end{figure*}

The execution of quantum circuits on quantum processors necessitates the mapping of logical qubits to physical qubits. Given that physical qubits are not fully connected, two-qubit (and multi-qubit) gates between non-adjacent qubits cannot be directly executed after mapping. This constraint requires the insertion of additional SWAP gates to facilitate the exchange of quantum states between qubits. As a SWAP gate is composed of three CNOT gates—a primary source of noise in the NISQ era—the introduction of an excessive number of SWAP gates can significantly degrade the fidelity of circuit execution. Consequently, minimizing the SWAP gate count is a paramount objective in quantum circuit compilation.
Fig.~\ref{fig:compiling} provides a representative example. For the depicted superconducting quantum topology, a two-qubit gate cannot be directly applied between physical qubits $Q_9$ and $Q_{10}$ due to the absence of a direct link. The initial compilation of the logical circuit in Fig.~\ref{fig:compiling}(b) would necessitate the insertion of three SWAP gates to satisfy connectivity. Subsequent qubit routing optimization, however, can reduce the required number of SWAP gates to a single instance, as demonstrated in Fig.~\ref{fig:compiling}(d). The final SWAP gate is typically omitted since it does not influence the measurement outcome.

\textbf{Problem Definition: }
The Qubit Mapping problem is defined as finding an optimal initial mapping of logical qubits to physical qubits and a corresponding qubit routing schedule. Given a quantum circuit, a quantum processor's coupling graph, and its noise conditions, the objective is to satisfy all two-qubit gate connectivity constraints by minimizing the overhead of auxiliary SWAP gates. This minimization is crucial for maximizing the overall circuit fidelity.

\subsection{Qubit Mapping and Routing Targets}

\subsubsection{Gate Count} 
In the NISQ era, quantum gate noise is non-negligible. To minimize overall circuit noise, reducing the number of quantum gates in the mapped circuit is essential. It should be noted that various studies~\cite{li2019tackling,wille2019mapping,niu2020hardware,molavi2022qubit,zulehner2018efficient,cowtan2019qubit,zhu2020dynamic,siraichi2018qubit,liu2022not,li2020qubit,zhou2020quantum,zhou2020monte,huang2022reinforcement,park2022fast,liu2021qucloud,niu2023enabling} may differ slightly in their approach to optimizing quantum gate count. These approaches can be broadly categorized into optimizing the total gate count, the number of added gates, SWAP gate count, and CNOT gate count. Notably, quantum gates added in the compiling stage are mainly the SWAP gates, and SWAP gates require only CNOT gates to implement. Thus, the objectives mentioned above are unified as optimizing gate count.


\subsubsection{Circuit Depth} 
This optimization objective reflects both the number of quantum gates and, to some extent, circuit execution time. Generally, deeper circuits imply longer execution time and increased likelihood of decoherence. Consequently, circuit depth is often included in the optimization objectives of quantum circuit mapping~\cite{li2019tackling,zulehner2018efficient,cowtan2019qubit,nannicini2022optimal,liu2022not,sinha2022qubit,pozzi2022using,liu2021qucloud}.

\subsubsection{Execution Time} 
A critical metric that correlates with both gate count and circuit depth. This is particularly important due to qubit decoherence, as a longer circuit runtime increases the probability of quantum states degrading, which leads to execution errors. Consequently, several studies~\cite{murali2019noise,niu2020hardware,zhang2021time} minimize the total execution time of the mapped circuit while satisfying the hardware's connectivity constraints. 

\subsubsection{Fidelity} 
A measure of similarity between quantum states is used to quantify the accuracy of circuit execution results by comparing the output state with the theoretically expected one. However, calculating fidelity directly on a quantum computer is challenging due to the complexity of measuring a complete qubit state. For this reason, many studies employ the Probability of a Successful Trial (PST) to reflect fidelity, as it can be directly computed from measurement outcomes \cite{tannu2019not,murali2019noise,niu2020hardware,liu2022not,liu2021qucloud,niu2023enabling}. Alternatively, \cite{ash2019qure} employed Kraus operators to model quantum gate noise and qubit decoherence, enabling direct calculation and optimization of the execution fidelity through classical simulation. 

\subsubsection{Turnaround Time}
The increasing demand for quantum resources has led to significant queueing delays on cloud platforms like those offered by IBM. To address this, multi-programming~\cite{das2019case} has been proposed as a way to execute multiple quantum circuits concurrently on a single quantum processor, thereby improving the utilization of limited quantum resources. This approach is further formulated as a scheduling problem \cite{wu2024reducing}. An important objective in this problem is to minimize the average turnaround time—the total time from the submission to the completion of a quantum job—given the quantum computer's coupling graph, noise conditions, and the queue of submitted jobs.

\subsection{Qubit Mapping and Routing Methods}
The quantum circuit mapping problem can be further divided into two sub-problems: initial mapping and routing. Initial mapping determines the one-to-one correspondence between logical and physical qubits at the circuit's outset, while routing ensures that two(multi)-qubit gates in the circuit satisfy connectivity constraints by inserting SWAP gates or employing other methods. Specific optimization approaches can be categorized as follows:

\subsubsection{Optimal Solution} 
\cite{siraichi2018qubit} presented a method to obtain an exact solution based on dynamic programming. Assuming a circuit with $n$ qubits and $m$ two-qubit gates, the solution to the subproblem is defined as $S(l,i)$, representing the minimum number of SWAP gates required to satisfy the constraints of the first $i$ two-qubit gates with the final qubit mapping being $l$. The solution to the original problem can thus be derived from various subproblem solutions. While this method yields the optimal solution, its time complexity of $\mathcal{O}(n!^2\cdot n\cdot m)$ significantly limits its scalability.

\subsubsection{Heuristic Algorithms} 
It generally plans SWAP gate paths using noise conditions and inter-qubit distances to find near-optimal solutions efficiently. As quantum circuits can be converted into DAGs, some studies apply classical graph algorithms to solve this problem. \cite{zulehner2018efficient} calculated local optimal solutions between circuit layers using the A$^*$ algorithm, incorporating a look-ahead mechanism to consider subsequent two-qubit gate impacts, approximating the global optimal solution. The authors in~\cite{zhang2021time} also employed the A$^*$ algorithm but consider individual gate execution time to optimize overall circuit execution time. \cite{tannu2019not} utilized the Dijkstra algorithm for an approximate solution.
To further reduce search space and improve solution quality, \cite{li2019tackling} introduced a scoring function guiding heuristic search, determined by the weighted sum of distances between all two-qubit gates in current and look-ahead layers. The heuristic search selects SWAP gates, minimizing this scoring function. Building on this, several studies~\cite{zhu2020dynamic,niu2020hardware,liu2022not,park2022fast,niu2023enabling} have proposed improvements. \cite{zhu2020dynamic} first selected SWAP gates that reduce current layer qubit distances as candidates, then chose those maximally reducing look-ahead layer qubit distances. \cite{niu2020hardware} incorporated noise and execution time terms into the distance function and introduced BRIDGE gates as SWAP alternatives. In~\cite{niu2023enabling}, they further considered SWAP and BRIDGE gate noise, enhancing scoring function realism. \cite{liu2022not} incorporated circuit optimization, reducing CNOT gate count through two-qubit block optimization and multiple two-qubit gate cancellation. \cite{park2022fast} added distance-based weights to each two-qubit gate term in the scoring function, assigning lower weights to longer distances. 

\subsubsection{Reduction} 
The quantum circuit mapping problem can be reduced to classical problems and solved using the corresponding solvers.  A series of works~\cite{murali2019noise,wille2019mapping,molavi2022qubit} transformed it into a Boolean Satisfiability Problem, while respectively employing SMT (Satisfiability Modulo Theories) solver, SAT solver, and MaxSAT solver to obtain solutions. Additionally, this problem can be reduced to a binary integer programming problem and solved using the commercial solver CPLEX~\cite{nannicini2022optimal}. As quantum circuits can be converted into DAGs, the initial mapping problem can be viewed as a Subgraph Isomorphism Problem. \cite{ash2019qure} first determined the graph structure of the initial mapping, then selected the least noisy subgraph from all isomorphic subgraphs on the quantum computer topology as the initial mapping. \cite{li2020qubit} considered the subgraph formed by all two-qubit gates in the top layer and found an isomorphic subgraph as the initial mapping. Due to the difficulty of achieving accurate subgraph matching, \cite{park2022fast} only required finding an approximation of the target subgraph structure as the initial mapping, while minimizing the difference between the two.

\subsubsection{Machine Learning} 
The quantum circuit mapping problem can be viewed as a decision problem, making it amenable to reinforcement learning approaches. \cite{pozzi2022using} defined the state as the current circuit layout and target mapping, actions as a series of parallel executable SWAP gates, and rewards determined by the number of plannable gates after action. They employ a simulated annealing algorithm for action search, with the action value function approximated by the Deep Q Learning algorithm. \cite{sinha2022qubit} augmented the state with unplanned gates and utilized the Monte Carlo tree search algorithm for action search, estimating the value function with a graph neural network. Reinforcement learning can also be applied exclusively to initial mapping. \cite{huang2022reinforcement} employed a multi-head attention mechanism to encode the circuit and decode it to obtain the initial mapping, then recursively extract sub-circuit structures and use the A$^*$ algorithm for routing. They use the number of gates added after circuit mapping as the reward. 

\section{Conclusion and Outlook}
This survey reviews methods related to quantum logic circuit design and compilation optimization, covering quantum circuit representation, synthesis, qubit mapping, and routing. To execute quantum programs on NISQ-era processors, algorithms must address constraints, e.g., limited gate sets, qubit coherence times, and noise. Efficiently transforming quantum algorithms into executable programs remains a critical challenge for demonstrating quantum supremacy. Over the past decade, advancements in artificial intelligence have significantly impacted quantum logic circuit design and optimization, yet existing AI algorithms address each step independently. When the process is divided into sub-modules handled by separate algorithms, achieving unified goals often leads to suboptimal performance. Therefore, future research should focus on end-to-end circuit synthesis optimization—from quantum algorithms to executable programs—by leveraging AI's ability to optimize the entire process holistically.

Quantum computing resources are currently accessible primarily through cloud-based platforms~\cite{huang2017homomorphic,huang2018demonstration}, with future deployment anticipated to shift towards distributed quantum computers~\cite{caleffi2024distributed}. Developing end-to-end algorithms on these platforms will enable seamless integration of quantum logic circuit design with real-time calibration data, facilitating the optimization of practical quantum algorithms. It will enhance quantum algorithm performance, reduce human intervention, and provide a novel paradigm for quantum computing implementation. As the quantum computing community advances towards FTQC, the scope of these optimization methods will extend beyond NISQ-era constraints, enabling the realization of fault-tolerant quantum algorithms with significantly improved scalability and robustness.

\begin{IEEEbiography}
[{\includegraphics[width=1in,height=1.25in,clip,keepaspectratio]{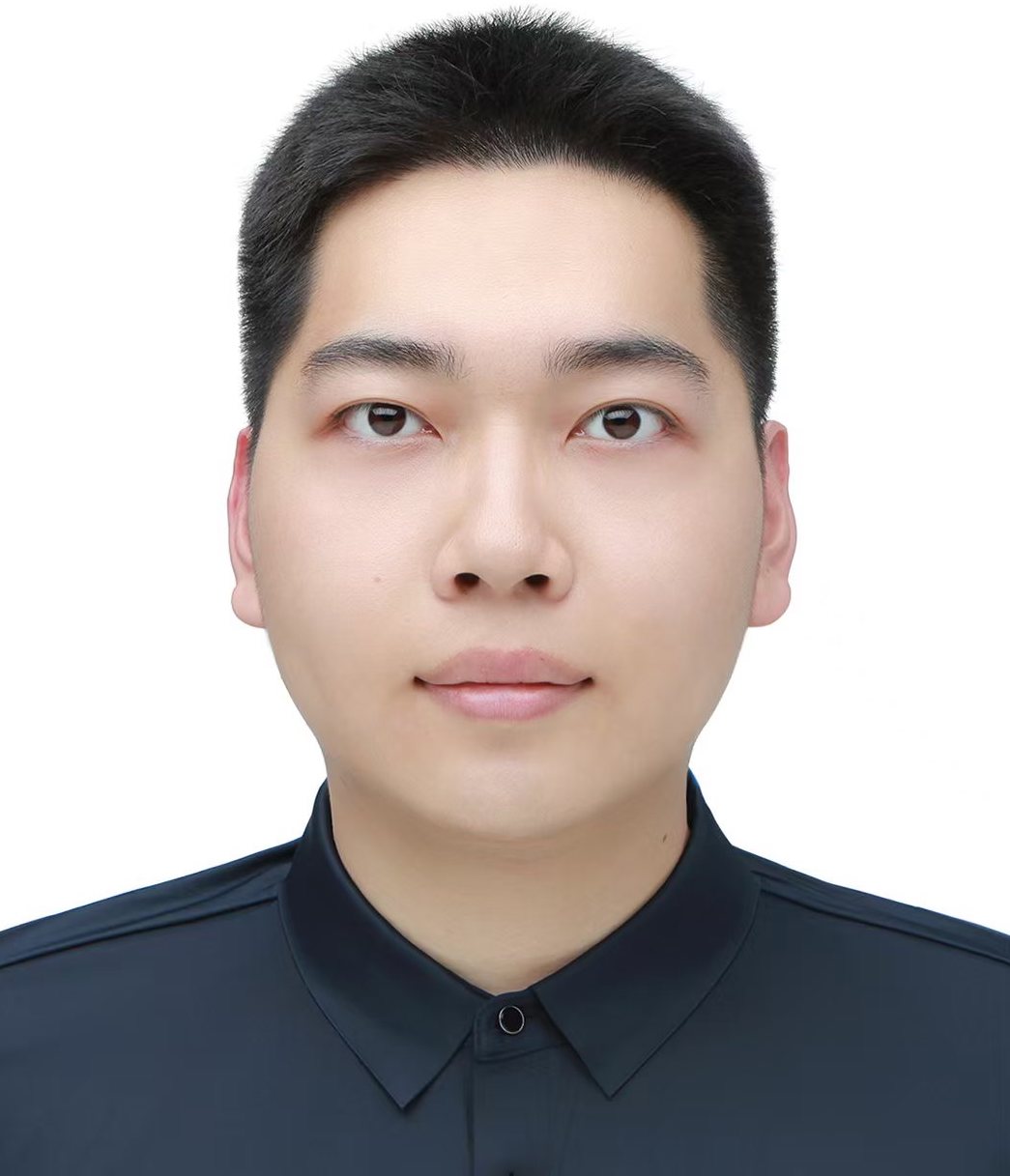}}] 
{Ge Yan} (M'25) received the PhD degree and B.S. (ACM Honored Program) degree both in Computer Science from Shanghai Jiao Tong University, Shanghai, China in 2025 and 2020, respectively. He is currently a postdoctoral researcher working in the lab led by Dr. Yuxuan Du and Professor Dacheng Tao at Nanyang Technological University, Singapore. His research interests are quantum AI and AI for quantum, with first-author papers in ICML, ICLR, NeurIPS, SIGKDD, etc.
\end{IEEEbiography}
\vspace{-30pt}
\begin{IEEEbiographynophoto}{Wenjie Wu}
received M.E. and B.E. in Computer Science and Automation from Shanghai Jiao Tong University in 2025 and 2022 respectively.
\end{IEEEbiographynophoto}
\vspace{-30pt}
\begin{IEEEbiographynophoto}{Yuheng Chen}
received B.E. in Computer Science from Shanghai Jiao Tong University in 2025 and currently a master student in computer science with the same university.
\end{IEEEbiographynophoto}
\vspace{-30pt}
\begin{IEEEbiographynophoto}{Kaisen Pan}
received B.S. in Computer Science from Shanghai Jiao Tong University in 2024. He is currently an independent researcher.
\end{IEEEbiographynophoto}
\vspace{-30pt}
\begin{IEEEbiographynophoto}{Xudong Lu}
received B.S. in Computer Science from Shanghai Jiao Tong University in 2023 and currently a PhD candidate with School of AI, in computer science of CUHK, Hongkong.
\end{IEEEbiographynophoto}
\vspace{-30pt}
\begin{IEEEbiographynophoto}{Zixiang Zhou}
He is currently an undergraduate student in electronic engineering, Shanghai Jiao Tong University, China.
\end{IEEEbiographynophoto}
\vspace{-30pt}
\begin{IEEEbiographynophoto}{Yuhan Wang}
He is currently an undergraduate student in physics, Shanghai Jiao Tong University, China. He will start the PhD program in Physics the same university in 2026.
\end{IEEEbiographynophoto}
\vspace{-30pt}
\begin{IEEEbiographynophoto}{Ruocheng Wang}
received B.S. in Physics from Shanghai Jiao Tong University in 2025 and currently a PhD candidate with School of AI, SJTU, China.
\end{IEEEbiographynophoto}
\vspace{-30pt}
\begin{IEEEbiography}
[{\includegraphics[width=1in,height=1.25in,clip,keepaspectratio]{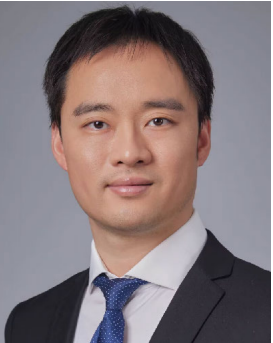}}] 
{Junchi Yan} (S'10-M'11-SM'21) is a Professor and Deputy Director with School of AI, Shanghai Jiao Tong University, Shanghai, China. Before that, he was a Research Staff Member with IBM and later an affiliated consultant Researcher with AWS AI Lab. His research interests are machine learning. He is the Associate Editor for IEEE TPAMI, Pattern Recognition. received IEEE CS AI'10 to Watch, IEEE CIS Early Career Award, CVPR Best Paper Candidate, ACL Outstanding Paper. He is a Fellow of IAPR, and on the board of ICML. He has published 20+ papers in top venues in quantum AI and AI for quantum.
\end{IEEEbiography}


\scriptsize
\begin{thebibliography}{100}

\bibitem{shor1994algorithms}
P.~W. Shor, ``Algorithms for quantum computation: discrete logarithms and factoring,'' in {\em Annual symposium on foundations of computer science}, pp.~124--134, 1994.

\bibitem{grover1996fast}
L.~K. Grover, ``A fast quantum mechanical algorithm for database search,'' in {\em ACM symposium on Theory of computing}, pp.~212--219, 1996.

\bibitem{huang2020superconducting}
H.-L. Huang, D.~Wu, D.~Fan, and X.~Zhu, ``Superconducting quantum computing: a review,'' {\em Science China Information Sciences}, vol.~63, no.~8, pp.~1--32, 2020.

\bibitem{preskill2018quantum}
J.~Preskill, ``Quantum computing in the nisq era and beyond,'' {\em Quantum}, vol.~2, p.~79, 2018.

\bibitem{arute2019quantum}
F.~Arute, K.~Arya, R.~Babbush, D.~Bacon, J.~C. Bardin, R.~Barends, R.~Biswas, S.~Boixo, F.~G. Brandao, D.~A. Buell, {\em et~al.}, ``Quantum supremacy using a programmable superconducting processor,'' {\em Nature}, vol.~574, no.~7779, pp.~505--510, 2019.

\bibitem{wu2021strong}
Y.~Wu, W.-S. Bao, S.~Cao, F.~Chen, M.-C. Chen, X.~Chen, T.-H. Chung, H.~Deng, Y.~Du, D.~Fan, {\em et~al.}, ``Strong quantum computational advantage using a superconducting quantum processor,'' {\em Physical review letters}, vol.~127, no.~18, p.~180501, 2021.

\bibitem{zhong2020quantum}
H.-S. Zhong, H.~Wang, Y.-H. Deng, M.-C. Chen, L.-C. Peng, Y.-H. Luo, J.~Qin, D.~Wu, X.~Ding, Y.~Hu, {\em et~al.}, ``Quantum computational advantage using photons,'' {\em Science}, vol.~370, no.~6523, pp.~1460--1463, 2020.

\bibitem{weinstein2001implementation}
Y.~S. Weinstein, M.~Pravia, E.~Fortunato, S.~Lloyd, and D.~G. Cory, ``Implementation of the quantum fourier transform,'' {\em Physical review letters}, vol.~86, no.~9, p.~1889, 2001.

\bibitem{huang2023near}
H.-L. Huang, X.-Y. Xu, C.~Guo, G.~Tian, S.-J. Wei, X.~Sun, W.-S. Bao, and G.-L. Long, ``Near-term quantum computing techniques: Variational quantum algorithms, error mitigation, circuit compilation, benchmarking and classical simulation,'' {\em Science China Physics, Mechanics \& Astronomy}, vol.~66, no.~5, p.~250302, 2023.

\bibitem{kusyk2021survey}
J.~Kusyk, S.~M. Saeed, and M.~U. Uyar, ``Survey on quantum circuit compilation for noisy intermediate-scale quantum computers: Artificial intelligence to heuristics,'' {\em IEEE Transactions on Quantum Engineering}, vol.~2, pp.~1--16, 2021.

\bibitem{deutsch1985quantum}
D.~Deutsch, ``Quantum theory, the church--turing principle and the universal quantum computer,'' {\em Proceedings of the Royal Society of London. A. Mathematical and Physical Sciences}, vol.~400, no.~1818, pp.~97--117, 1985.

\bibitem{maslov2008quantum}
D.~Maslov, G.~W. Dueck, D.~M. Miller, and C.~Negrevergne, ``Quantum circuit simplification and level compaction,'' {\em IEEE Transactions on Computer-Aided Design of Integrated Circuits and Systems}, vol.~27, no.~3, pp.~436--444, 2008.

\bibitem{amy2014polynomial}
M.~Amy, D.~Maslov, and M.~Mosca, ``Polynomial-time t-depth optimization of clifford+ t circuits via matroid partitioning,'' {\em IEEE Transactions on Computer-Aided Design of Integrated Circuits and Systems}, vol.~33, no.~10, pp.~1476--1489, 2014.

\bibitem{coecke2008interacting}
B.~Coecke and R.~Duncan, ``Interacting quantum observables,'' in {\em International Colloquium on Automata, Languages, and Programming}, pp.~298--310, 2008.

\bibitem{biamonte2017tensor}
J.~Biamonte and V.~Bergholm, ``Tensor networks in a nutshell,'' {\em arXiv preprint arXiv:1708.00006}, 2017.

\bibitem{bijwe2022implementing}
S.~Bijwe, A.~K. Chauhan, and S.~K. Sanadhya, ``Implementing grover oracle for lightweight block ciphers under depth constraints,'' in {\em Australasian Conference on Information Security and Privacy}, pp.~85--105, 2022.

\bibitem{rahman2022grover}
M.~Rahman and G.~Paul, ``Grover on katan: Quantum resource estimation,'' {\em IEEE Transactions on Quantum Engineering}, vol.~3, pp.~1--9, 2022.

\bibitem{grimsley2019adaptive}
H.~R. Grimsley, S.~E. Economou, E.~Barnes, and N.~J. Mayhall, ``An adaptive variational algorithm for exact molecular simulations on a quantum computer,'' {\em Nature communications}, vol.~10, no.~1, p.~3007, 2019.

\bibitem{du2022quantum}
Y.~Du, T.~Huang, S.~You, M.-H. Hsieh, and D.~Tao, ``Quantum circuit architecture search for variational quantum algorithms,'' {\em npj Quantum Information}, vol.~8, no.~1, pp.~1--8, 2022.

\bibitem{wu2023quantumdarts}
W.~Wu, G.~Yan, X.~Lu, K.~Pan, and J.~Yan, ``Quantumdarts: differentiable quantum architecture search for variational quantum algorithms,'' in {\em ICML}, pp.~37745--37764, 2023.

\bibitem{chen2019machine}
H.~Chen, M.~Vasmer, N.~P. Breuckmann, and E.~Grant, ``Machine learning logical gates for quantum error correction,'' {\em arXiv preprint arXiv:1912.10063}, 2019.

\bibitem{cong2019quantum}
I.~Cong, S.~Choi, and M.~D. Lukin, ``Quantum convolutional neural networks,'' {\em Nature Physics}, vol.~15, no.~12, pp.~1273--1278, 2019.

\bibitem{rattew2023non}
A.~G. Rattew and P.~Rebentrost, ``Non-linear transformations of quantum amplitudes: Exponential improvement, generalization, and applications,'' {\em arXiv preprint arXiv:2309.09839}, 2023.

\bibitem{guo2024nonlinear}
N.~Guo, K.~Mitarai, and K.~Fujii, ``Nonlinear transformation of complex amplitudes via quantum singular value transformation,'' {\em Physical Review Research}, vol.~6, no.~4, p.~043227, 2024.

\bibitem{mckeeman1965peephole}
W.~M. McKeeman, ``Peephole optimization,'' {\em Communications of the ACM}, vol.~8, no.~7, pp.~443--444, 1965.

\bibitem{heyfron2018efficient}
L.~E. Heyfron and E.~T. Campbell, ``An efficient quantum compiler that reduces t count,'' {\em Quantum Science and Technology}, vol.~4, no.~1, p.~015004, 2018.

\bibitem{amy2019t}
M.~Amy and M.~Mosca, ``T-count optimization and reed--muller codes,'' {\em IEEE Transactions on Information Theory}, vol.~65, no.~8, pp.~4771--4784, 2019.

\bibitem{schneider2023sat}
S.~Schneider, L.~Burgholzer, and R.~Wille, ``A sat encoding for optimal clifford circuit synthesis,'' in {\em Asia and South Pacific Design Automation Conference}, pp.~190--195, 2023.

\bibitem{fosel2021quantum}
T.~F{\"o}sel, M.~Y. Niu, F.~Marquardt, and L.~Li, ``Quantum circuit optimization with deep reinforcement learning,'' {\em arXiv preprint arXiv:2103.07585}, 2021.

\bibitem{li2023quarl}
Z.~Li, J.~Peng, Y.~Mei, S.~Lin, Y.~Wu, O.~Padon, and Z.~Jia, ``Quarl: A learning-based quantum circuit optimizer,'' {\em Proceedings of the ACM on Programming Languages}, vol.~8, no.~OOPSLA1, pp.~555--582, 2024.

\bibitem{riu2023reinforcement}
J.~Riu, J.~Nogu{\'e}, G.~Vilaplana, A.~Garcia-Saez, and M.~P. Estarellas, ``Reinforcement learning based quantum circuit optimization via zx-calculus,'' {\em Quantum}, vol.~9, p.~1758, 2025.

\bibitem{tang2021qubit}
H.~L. Tang, V.~Shkolnikov, G.~S. Barron, H.~R. Grimsley, N.~J. Mayhall, E.~Barnes, and S.~E. Economou, ``qubit-adapt-vqe: An adaptive algorithm for constructing hardware-efficient ans{\"a}tze on a quantum processor,'' {\em PRX Quantum}, vol.~2, no.~2, p.~020310, 2021.

\bibitem{yordanov2021qubit}
Y.~S. Yordanov, V.~Armaos, C.~H. Barnes, and D.~R. Arvidsson-Shukur, ``Qubit-excitation-based adaptive variational quantum eigensolver,'' {\em Communications Physics}, vol.~4, no.~1, p.~228, 2021.

\bibitem{magoulas2023linear}
I.~Magoulas and F.~A. Evangelista, ``Linear-scaling quantum circuits for computational chemistry,'' {\em Journal of Chemical Theory and Computation}, vol.~19, no.~15, pp.~4815--4821, 2023.

\bibitem{tannu2019not}
S.~S. Tannu and M.~K. Qureshi, ``Not all qubits are created equal: A case for variability-aware policies for nisq-era quantum computers,'' in {\em Proceedings of the Twenty-Fourth International Conference on Architectural Support for Programming Languages and Operating Systems}, pp.~987--999, 2019.

\bibitem{murali2019noise}
P.~Murali, J.~M. Baker, A.~Javadi-Abhari, F.~T. Chong, and M.~Martonosi, ``Noise-adaptive compiler mappings for noisy intermediate-scale quantum computers,'' in {\em PLOS}, pp.~1015--1029, 2019.

\bibitem{niu2020hardware}
S.~Niu, A.~Suau, G.~Staffelbach, and A.~Todri-Sanial, ``A hardware-aware heuristic for the qubit mapping problem in the nisq era,'' {\em IEEE Transactions on Quantum Engineering}, vol.~1, pp.~1--14, 2020.

\bibitem{liu2022not}
J.~Liu, P.~Li, and H.~Zhou, ``Not all swaps have the same cost: A case for optimization-aware qubit routing,'' in {\em HPCA}, pp.~709--725, 2022.

\bibitem{liu2021qucloud}
L.~Liu and X.~Dou, ``Qucloud: A new qubit mapping mechanism for multi-programming quantum computing in cloud environment,'' in {\em HPCA}, pp.~167--178, 2021.

\bibitem{niu2023enabling}
S.~Niu and A.~Todri-Sanial, ``Enabling multi-programming mechanism for quantum computing in the nisq era,'' {\em Quantum}, vol.~7, p.~925, 2023.

\bibitem{siraichi2018qubit}
M.~Y. Siraichi, V.~F.~d. Santos, C.~Collange, and F.~M.~Q. Pereira, ``Qubit allocation,'' in {\em International Symposium on Code Generation and Optimization}, pp.~113--125, 2018.

\bibitem{zulehner2018efficient}
A.~Zulehner, A.~Paler, and R.~Wille, ``An efficient methodology for mapping quantum circuits to the ibm qx architectures,'' {\em IEEE Transactions on Computer-Aided Design of Integrated Circuits and Systems}, vol.~38, no.~7, pp.~1226--1236, 2018.

\bibitem{wille2019mapping}
R.~Wille, L.~Burgholzer, and A.~Zulehner, ``Mapping quantum circuits to ibm qx architectures using the minimal number of swap and h operations,'' in {\em Proceedings of the 56th Annual Design Automation Conference 2019}, pp.~1--6, 2019.

\bibitem{molavi2022qubit}
A.~Molavi, A.~Xu, M.~Diges, L.~Pick, S.~Tannu, and A.~Albarghouthi, ``Qubit mapping and routing via maxsat,'' in {\em MICRO}, pp.~1078--1091, 2022.

\bibitem{pozzi2022using}
M.~G. Pozzi, S.~J. Herbert, A.~Sengupta, and R.~D. Mullins, ``Using reinforcement learning to perform qubit routing in quantum compilers,'' {\em ACM Transactions on Quantum Computing}, vol.~3, no.~2, pp.~1--25, 2022.

\bibitem{sinha2022qubit}
A.~Sinha, U.~Azad, and H.~Singh, ``Qubit routing using graph neural network aided monte carlo tree search,'' in {\em AAAI}, vol.~36, pp.~9935--9943, 2022.

\bibitem{huang2022reinforcement}
C.-Y. Huang, C.-H. Lien, and W.-K. Mak, ``Reinforcement learning and dear framework for solving the qubit mapping problem,'' in {\em ICCAD}, pp.~1--9, 2022.

\bibitem{nielsen_chuang_2010}
M.~A. Nielsen and I.~L. Chuang, {\em Quantum Computation and Quantum Information: 10th Anniversary Edition}.
\newblock Cambridge University Press, 2010.

\bibitem{forest2015exact}
S.~Forest, D.~Gosset, V.~Kliuchnikov, and D.~McKinnon, ``Exact synthesis of single-qubit unitaries over clifford-cyclotomic gate sets,'' {\em Journal of Mathematical Physics}, vol.~56, no.~8, 2015.

\bibitem{gottesman1997stabilizer}
D.~Gottesman, {\em Stabilizer codes and quantum error correction}.
\newblock California Institute of Technology, 1997.

\bibitem{gottesman1998theory}
D.~Gottesman, ``Theory of fault-tolerant quantum computation,'' {\em Physical Review A}, vol.~57, no.~1, p.~127, 1998.

\bibitem{gottesman1998heisenberg}
D.~Gottesman, ``The heisenberg representation of quantum computers,'' {\em arXiv preprint quant-ph/9807006}, 1998.

\bibitem{miller2011elementary}
D.~M. Miller, R.~Wille, and Z.~Sasanian, ``Elementary quantum gate realizations for multiple-control toffoli gates,'' in {\em International symposium on multiple-valued logic}, pp.~288--293, 2011.

\bibitem{maslov2003simplification}
D.~Maslov, G.~W. Dueck, and D.~M. Miller, ``Simplification of toffoli networks via templates,'' in {\em Symposium on Integrated Circuits and Systems Design}, pp.~53--58, 2003.

\bibitem{maslov2005quantum}
D.~Maslov, C.~Young, D.~M. Miller, and G.~W. Dueck, ``Quantum circuit simplification using templates,'' in {\em Design, Automation and Test in Europe}, pp.~1208--1213, 2005.

\bibitem{stanisic2022observing}
S.~Stanisic, J.~L. Bosse, F.~M. Gambetta, R.~A. Santos, W.~Mruczkiewicz, T.~E. O’Brien, E.~Ostby, and A.~Montanaro, ``Observing ground-state properties of the fermi-hubbard model using a scalable algorithm on a quantum computer,'' {\em Nature communications}, vol.~13, no.~1, p.~5743, 2022.

\bibitem{he2023gsqas}
Z.~He, M.~Deng, S.~Zheng, L.~Li, and H.~Situ, ``Gsqas: graph self-supervised quantum architecture search,'' {\em Physica A: Statistical Mechanics and its Applications}, vol.~630, p.~129286, 2023.

\bibitem{he2023gnn}
Z.~He, X.~Zhang, C.~Chen, Z.~Huang, Y.~Zhou, and H.~Situ, ``A gnn-based predictor for quantum architecture search,'' {\em Quantum Information Processing}, vol.~22, no.~2, p.~128, 2023.

\bibitem{li2019tackling}
G.~Li, Y.~Ding, and Y.~Xie, ``Tackling the qubit mapping problem for nisq-era quantum devices,'' in {\em International Conference on Architectural Support for Programming Languages and Operating Systems}, pp.~1001--1014, 2019.

\bibitem{beaudoin2024altgraph}
C.~Beaudoin, K.~Phalak, and S.~Ghosh, ``Altgraph: Redesigning quantum circuits using generative graph models for efficient optimization,'' in {\em Proceedings of the Great Lakes Symposium on VLSI 2024}, pp.~44--49, 2024.

\bibitem{niemann2015qmdds}
P.~Niemann, R.~Wille, D.~M. Miller, M.~A. Thornton, and R.~Drechsler, ``Qmdds: Efficient quantum function representation and manipulation,'' {\em IEEE Transactions on Computer-Aided Design of Integrated Circuits and Systems}, vol.~35, no.~1, pp.~86--99, 2015.

\bibitem{amy2013meet}
M.~Amy, D.~Maslov, M.~Mosca, and M.~Roetteler, ``A meet-in-the-middle algorithm for fast synthesis of depth-optimal quantum circuits,'' {\em IEEE Transactions on Computer-Aided Design of Integrated Circuits and Systems}, vol.~32, no.~6, pp.~818--830, 2013.

\bibitem{nam2018automated}
Y.~Nam, N.~J. Ross, Y.~Su, A.~M. Childs, and D.~Maslov, ``Automated optimization of large quantum circuits with continuous parameters,'' {\em npj Quantum Information}, vol.~4, no.~1, p.~23, 2018.

\bibitem{meuli2018sat}
G.~Meuli, M.~Soeken, and G.~De~Micheli, ``Sat-based $\{$CNOT, T$\}$ quantum circuit synthesis,'' in {\em Reversible Computation: 10th International Conference, RC 2018, Leicester, UK, September 12-14, 2018, Proceedings 10}, pp.~175--188, 2018.

\bibitem{amy2018controlled}
M.~Amy, P.~Azimzadeh, and M.~Mosca, ``On the controlled-not complexity of controlled-not--phase circuits,'' {\em Quantum Science and Technology}, vol.~4, no.~1, p.~015002, 2018.

\bibitem{vandaele2022phase}
V.~Vandaele, S.~Martiel, and T.~G. de~Brugi{\`e}re, ``Phase polynomials synthesis algorithms for nisq architectures and beyond,'' {\em Quantum Science and Technology}, vol.~7, no.~4, p.~045027, 2022.

\bibitem{ruiz2025quantum}
F.~J. Ruiz, T.~Laakkonen, J.~Bausch, M.~Balog, M.~Barekatain, F.~J. Heras, A.~Novikov, N.~Fitzpatrick, B.~Romera-Paredes, J.~van~de Wetering, {\em et~al.}, ``Quantum circuit optimization with alphatensor,'' {\em Nature Machine Intelligence}, pp.~1--12, 2025.

\bibitem{penrose1971applications}
R.~Penrose, ``Applications of negative dimensional tensors,'' {\em Combinatorial mathematics and its applications}, vol.~1, pp.~221--244, 1971.

\bibitem{orus2014practical}
R.~Or{\'u}s, ``A practical introduction to tensor networks: Matrix product states and projected entangled pair states,'' {\em Annals of physics}, vol.~349, pp.~117--158, 2014.

\bibitem{markov2008simulating}
I.~L. Markov and Y.~Shi, ``Simulating quantum computation by contracting tensor networks,'' {\em SIAM Journal on Computing}, vol.~38, no.~3, pp.~963--981, 2008.

\bibitem{chi1997optimizing}
L.~Chi-Chung, P.~Sadayappan, and R.~Wenger, ``On optimizing a class of multi-dimensional loops with reduction for parallel execution,'' {\em Parallel Processing Letters}, vol.~7, no.~02, pp.~157--168, 1997.

\bibitem{schutski2020simple}
R.~Schutski, T.~Khakhulin, I.~Oseledets, and D.~Kolmakov, ``Simple heuristics for efficient parallel tensor contraction and quantum circuit simulation,'' {\em Physical Review A}, vol.~102, no.~6, p.~062614, 2020.

\bibitem{huang2021efficient}
C.~Huang, F.~Zhang, M.~Newman, X.~Ni, D.~Ding, J.~Cai, X.~Gao, T.~Wang, F.~Wu, G.~Zhang, {\em et~al.}, ``Efficient parallelization of tensor network contraction for simulating quantum computation,'' {\em Nature Computational Science}, vol.~1, no.~9, pp.~578--587, 2021.

\bibitem{lyakh2015efficient}
D.~I. Lyakh, ``An efficient tensor transpose algorithm for multicore cpu, intel xeon phi, and nvidia tesla gpu,'' {\em Computer Physics Communications}, vol.~189, pp.~84--91, 2015.

\bibitem{vincent2022jet}
T.~Vincent, L.~J. O'Riordan, M.~Andrenkov, J.~Brown, N.~Killoran, H.~Qi, and I.~Dhand, ``Jet: Fast quantum circuit simulations with parallel task-based tensor-network contraction,'' {\em Quantum}, vol.~6, p.~709, 2022.

\bibitem{huggins2019towards}
W.~Huggins, P.~Patil, B.~Mitchell, K.~B. Whaley, and E.~M. Stoudenmire, ``Towards quantum machine learning with tensor networks,'' {\em Quantum Science and technology}, vol.~4, no.~2, p.~024001, 2019.

\bibitem{haghshenas2022variational}
R.~Haghshenas, J.~Gray, A.~C. Potter, and G.~K.-L. Chan, ``Variational power of quantum circuit tensor networks,'' {\em Physical Review X}, vol.~12, no.~1, p.~011047, 2022.

\bibitem{Kissinger_2020}
A.~Kissinger and J.~van~de Wetering, ``{PyZX}: Large scale automated diagrammatic reasoning,'' {\em Electronic Proceedings in Theoretical Computer Science}, vol.~318, pp.~229--241, may 2020.

\bibitem{yeung2020diagrammatic}
R.~Yeung, ``Diagrammatic design and study of ans{\"a}tze for quantum machine learning,'' {\em arXiv preprint arXiv:2011.11073}, 2020.

\bibitem{Duncan_2020}
R.~Duncan, A.~Kissinger, S.~Perdrix, and J.~van~de Wetering, ``Graph-theoretic simplification of quantum circuits with the {ZX}-calculus,'' {\em Quantum}, vol.~4, p.~279, jun 2020.

\bibitem{kissinger2015quantomatic}
A.~Kissinger and V.~Zamdzhiev, ``Quantomatic: A proof assistant for diagrammatic reasoning,'' in {\em International Conference on Automated Deduction}, pp.~326--336, 2015.

\bibitem{Fagan_2019}
A.~Fagan and R.~Duncan, ``Optimising clifford circuits with quantomatic,'' {\em Electronic Proceedings in Theoretical Computer Science}, vol.~287, pp.~85--105, jan 2019.

\bibitem{van_de_Wetering_2021}
J.~van~de Wetering, ``Constructing quantum circuits with global gates,'' {\em New Journal of Physics}, vol.~23, p.~043015, apr 2021.

\bibitem{de_Beaudrap_2020}
N.~de~Beaudrap, X.~Bian, and Q.~Wang, ``Techniques to reduce $\pi$/4-parity-phase circuits, motivated by the {ZX} calculus,'' {\em Electronic Proceedings in Theoretical Computer Science}, vol.~318, pp.~131--149, may 2020.

\bibitem{Munson_2021}
A.~Munson, B.~Coecke, and Q.~Wang, ``{AND}-gates in {ZX}-calculus: Spider nest identities and {QBC}-completeness,'' {\em Electronic Proceedings in Theoretical Computer Science}, vol.~340, pp.~230--255, sep 2021.

\bibitem{Kissinger_2020_2}
A.~Kissinger and J.~van~de Wetering, ``Reducing the number of non-clifford gates in quantum circuits,'' {\em Physical Review A}, vol.~102, aug 2020.

\bibitem{gogioso2022annealing}
S.~Gogioso and R.~Yeung, ``Annealing optimisation of mixed zx phase circuits,'' {\em arXiv preprint arXiv:2206.11839}, 2022.

\bibitem{kissinger2019cnot}
A.~Kissinger and A.~M.-v. de~Griend, ``Cnot circuit extraction for topologically-constrained quantum memories,'' {\em arXiv preprint arXiv:1904.00633}, 2019.

\bibitem{jaques2020implementing}
S.~Jaques, M.~Naehrig, M.~Roetteler, and F.~Virdia, ``Implementing grover oracles for quantum key search on aes and lowmc,'' in {\em Advances in Cryptology--EUROCRYPT 2020: 39th Annual International Conference on the Theory and Applications of Cryptographic Techniques, Zagreb, Croatia, May 10--14, 2020, Proceedings, Part II 30}, pp.~280--310, 2020.

\bibitem{kitaev2002classical}
A.~Y. Kitaev, A.~Shen, and M.~N. Vyalyi, {\em Classical and quantum computation}.
\newblock No.~47, American Mathematical Soc., 2002.

\bibitem{vartiainen2004efficient}
J.~J. Vartiainen, M.~M{\"o}tt{\"o}nen, and M.~M. Salomaa, ``Efficient decomposition of quantum gates,'' {\em Physical review letters}, vol.~92, no.~17, p.~177902, 2004.

\bibitem{dawson2005solovay}
C.~M. Dawson and M.~A. Nielsen, ``The solovay-kitaev algorithm,'' {\em arXiv preprint quant-ph/0505030}, 2005.

\bibitem{zhang2022differentiable}
S.~Zhang, C.~Hsieh, S.~Zhang, and H.~Yao, ``Differentiable quantum architecture search,'' {\em Quantum Science and Technology}, vol.~7, no.~4, p.~045023, 2022.

\bibitem{lu2023qas}
X.~Lu, K.~Pan, G.~Yan, J.~Shan, W.~Wu, and J.~Yan, ``Qas-bench: rethinking quantum architecture search and a benchmark,'' in {\em ICML}, pp.~22880--22898, 2023.

\bibitem{liu2018darts}
H.~Liu, K.~Simonyan, and Y.~Yang, ``Darts: Differentiable architecture search,'' {\em arXiv preprint arXiv:1806.09055}, 2018.

\bibitem{WangECCV22}
X.~Wang, J.~Lin, J.~Zhao, X.~Yang, and J.~Yan, ``Eautodet: Efficient architecture search for object detection,'' in {\em European Conference on Computer Vision}, 2022.

\bibitem{ashhab2022numerical}
S.~Ashhab, N.~Yamamoto, F.~Yoshihara, and K.~Semba, ``Numerical analysis of quantum circuits for state preparation and unitary operator synthesis,'' {\em Physical Review A}, vol.~106, no.~2, p.~022426, 2022.

\bibitem{ashhab2024quantum}
S.~Ashhab, F.~Yoshihara, M.~Tsuji, M.~Sato, and K.~Semba, ``Quantum circuit synthesis via a random combinatorial search,'' {\em Physical Review A}, vol.~109, no.~5, p.~052605, 2024.

\bibitem{ding2006evolving}
S.~Ding, Z.~Jin, and Q.~Yang, ``Evolving quantum oracles with hybrid quantum-inspired evolutionary algorithm,'' {\em arXiv preprint quant-ph/0610105}, 2006.

\bibitem{li2017approximate}
R.~Li, U.~Alvarez-Rodriguez, L.~Lamata, and E.~Solano, ``Approximate quantum adders with genetic algorithms: an ibm quantum experience,'' {\em Quantum Measurements and Quantum Metrology}, vol.~4, no.~1, pp.~1--7, 2017.

\bibitem{deibuk2015design}
V.~G. Deibuk and A.~V. Biloshytskyi, ``Design of a ternary reversible/quantum adder using genetic algorithm,'' {\em International Journal of Information Technology and Computer Science}, vol.~7, no.~9, pp.~38--45, 2015.

\bibitem{lamata2018quantum}
L.~Lamata, U.~Alvarez-Rodriguez, J.~D. Mart{\'\i}n-Guerrero, M.~Sanz, and E.~Solano, ``Quantum autoencoders via quantum adders with genetic algorithms,'' {\em Quantum Science and Technology}, vol.~4, no.~1, p.~014007, 2018.

\bibitem{peruzzo2014variational}
A.~Peruzzo, J.~McClean, P.~Shadbolt, M.-H. Yung, X.-Q. Zhou, P.~J. Love, A.~Aspuru-Guzik, and J.~L. O’brien, ``A variational eigenvalue solver on a photonic quantum processor,'' {\em Nature communications}, vol.~5, no.~1, p.~4213, 2014.

\bibitem{berazin2012method}
F.~Berazin, {\em The method of second quantization}, vol.~24.
\newblock Elsevier, 2012.

\bibitem{jordan1993paulische}
P.~Jordan and E.~Wigner, ``{\"U}ber das paulische {\"a}quivalenzverbot,'' {\em Zeitschrift f{\"u}r Physik}, vol.~47, no.~9, pp.~631--651, 1928.

\bibitem{taube2006new}
A.~G. Taube and R.~J. Bartlett, ``New perspectives on unitary coupled-cluster theory,'' {\em International journal of quantum chemistry}, vol.~106, no.~15, pp.~3393--3401, 2006.

\bibitem{barkoutsos2018quantum}
P.~K. Barkoutsos, J.~F. Gonthier, I.~Sokolov, N.~Moll, G.~Salis, A.~Fuhrer, M.~Ganzhorn, D.~J. Egger, M.~Troyer, A.~Mezzacapo, {\em et~al.}, ``Quantum algorithms for electronic structure calculations: Particle-hole hamiltonian and optimized wave-function expansions,'' {\em Physical Review A}, vol.~98, no.~2, p.~022322, 2018.

\bibitem{sapova2022variational}
M.~D. Sapova and A.~K. Fedorov, ``Variational quantum eigensolver techniques for simulating carbon monoxide oxidation,'' {\em Communications Physics}, vol.~5, no.~1, pp.~1--13, 2022.

\bibitem{ostaszewski2021reinforcement}
M.~Ostaszewski, L.~M. Trenkwalder, W.~Masarczyk, E.~Scerri, and V.~Dunjko, ``Reinforcement learning for optimization of variational quantum circuit architectures,'' {\em NeurIPS}, vol.~34, pp.~18182--18194, 2021.

\bibitem{wang2022quantumnas}
H.~Wang, Y.~Ding, J.~Gu, Y.~Lin, D.~Z. Pan, F.~T. Chong, and S.~Han, ``Quantumnas: Noise-adaptive search for robust quantum circuits,'' in {\em HPCA}, pp.~692--708, 2022.

\bibitem{farhi2014quantum}
E.~Farhi, J.~Goldstone, and S.~Gutmann, ``A quantum approximate optimization algorithm,'' {\em arXiv preprint arXiv:1411.4028}, 2014.

\bibitem{majumdar2021depth}
R.~Majumdar, D.~Bhoumik, D.~Madan, D.~Vinayagamurthy, S.~Raghunathan, and S.~Sur-Kolay, ``Depth optimized ansatz circuit in qaoa for max-cut,'' {\em arXiv preprint arXiv:2110.04637}, 2021.

\bibitem{majumdar2021optimizing}
R.~Majumdar, D.~Madan, D.~Bhoumik, D.~Vinayagamurthy, S.~Raghunathan, and S.~Sur-Kolay, ``Optimizing ansatz design in qaoa for max-cut,'' {\em arXiv preprint arXiv:2106.02812}, 2021.

\bibitem{duong2022quantum}
T.~Duong, S.~T. Truong, M.~Tam, B.~Bach, J.-Y. Ryu, and J.-K.~K. Rhee, ``Quantum neural architecture search with quantum circuits metric and bayesian optimization,'' {\em arXiv preprint arXiv:2206.14115}, 2022.

\bibitem{lecun2015deep}
Y.~LeCun, Y.~Bengio, and G.~Hinton, ``Deep learning,'' {\em nature}, vol.~521, no.~7553, pp.~436--444, 2015.

\bibitem{gu2018recent}
J.~Gu, Z.~Wang, J.~Kuen, L.~Ma, A.~Shahroudy, B.~Shuai, T.~Liu, X.~Wang, G.~Wang, J.~Cai, {\em et~al.}, ``Recent advances in convolutional neural networks,'' {\em Pattern recognition}, vol.~77, pp.~354--377, 2018.

\bibitem{bausch2020recurrent}
J.~Bausch, ``Recurrent quantum neural networks,'' {\em Advances in neural information processing systems}, vol.~33, pp.~1368--1379, 2020.

\bibitem{huang2022quantum}
H.-Y. Huang, M.~Broughton, J.~Cotler, S.~Chen, J.~Li, M.~Mohseni, H.~Neven, R.~Babbush, R.~Kueng, J.~Preskill, {\em et~al.}, ``Quantum advantage in learning from experiments,'' {\em Science}, vol.~376, no.~6598, pp.~1182--1186, 2022.

\bibitem{zhang2021neural}
S.-X. Zhang, C.-Y. Hsieh, S.~Zhang, and H.~Yao, ``Neural predictor based quantum architecture search,'' {\em Machine Learning: Science and Technology}, vol.~2, no.~4, p.~045027, 2021.

\bibitem{cai2021bosonic}
W.~Cai, Y.~Ma, W.~Wang, C.-L. Zou, and L.~Sun, ``Bosonic quantum error correction codes in superconducting quantum circuits,'' {\em Fundamental Research}, vol.~1, no.~1, pp.~50--67, 2021.

\bibitem{rigby2021heuristics}
A.~Rigby, {\em Heuristics in quantum error correction}.
\newblock PhD thesis, University of Tasmania, 2021.

\bibitem{nautrup2019optimizing}
H.~P. Nautrup, N.~Delfosse, V.~Dunjko, H.~J. Briegel, and N.~Friis, ``Optimizing quantum error correction codes with reinforcement learning,'' {\em Quantum}, vol.~3, p.~215, 2019.

\bibitem{zeng2022approximate}
Y.~Zeng, Z.-Y. Zhou, E.~Rinaldi, C.~Gneiting, and F.~Nori, ``Approximate autonomous quantum error correction with reinforcement learning,'' {\em Physical Review Letters}, vol.~131, no.~5, p.~050601, 2023.

\bibitem{schuld2014quest}
M.~Schuld, I.~Sinayskiy, and F.~Petruccione, ``The quest for a quantum neural network,'' {\em Quantum Information Processing}, vol.~13, pp.~2567--2586, 2014.

\bibitem{williams1999automated}
C.~P. Williams and A.~G. Gray, ``Automated design of quantum circuits,'' in {\em NASA International Conference on Quantum Computing and Quantum Communications}, pp.~113--125, 1999.

\bibitem{khatri2019quantum}
S.~Khatri, R.~LaRose, A.~Poremba, L.~Cincio, A.~T. Sornborger, and P.~J. Coles, ``Quantum-assisted quantum compiling,'' {\em Quantum}, vol.~3, p.~140, 2019.

\bibitem{ye2021quantum}
E.~Ye and S.~Y.-C. Chen, ``Quantum architecture search via continual reinforcement learning,'' {\em arXiv preprint arXiv:2112.05779}, 2021.

\bibitem{kuo2021quantum}
E.-J. Kuo, Y.-L.~L. Fang, and S.~Y.-C. Chen, ``Quantum architecture search via deep reinforcement learning,'' {\em arXiv preprint arXiv:2104.07715}, 2021.

\bibitem{patel2024curriculum}
Y.~J. Patel, A.~Kundu, M.~Ostaszewski, X.~Bonet-Monroig, V.~Dunjko, and O.~Danaci, ``Curriculum reinforcement learning for quantum architecture search under hardware errors,'' {\em arXiv preprint arXiv:2402.03500}, 2024.

\bibitem{he2024training}
Z.~He, M.~Deng, S.~Zheng, L.~Li, and H.~Situ, ``Training-free quantum architecture search,'' in {\em AAAI}, vol.~38, pp.~12430--12438, 2024.

\bibitem{massey2004evolving}
P.~Massey, J.~A. Clark, and S.~Stepney, ``Evolving quantum circuits and programs through genetic programming,'' in {\em Genetic and Evolutionary Computation Conference}, pp.~569--580, 2004.

\bibitem{bang2014genetic}
J.~Bang and S.~Yoo, ``A genetic-algorithm-based method to find unitary transformations for any desired quantum computation and application to a one-bit oracle decision problem,'' {\em Journal of the Korean Physical Society}, vol.~65, no.~12, pp.~2001--2008, 2014.

\bibitem{las2016genetic}
U.~Las~Heras, U.~Alvarez-Rodriguez, E.~Solano, and M.~Sanz, ``Genetic algorithms for digital quantum simulations,'' {\em Physical review letters}, vol.~116, no.~23, p.~230504, 2016.

\bibitem{potovcek2018multi}
V.~Poto{\v{c}}ek, A.~P. Reynolds, A.~Fedrizzi, and D.~W. Corne, ``Multi-objective evolutionary algorithms for quantum circuit discovery,'' {\em arXiv preprint arXiv:1812.04458}, 2018.

\bibitem{Diag1}
A.~Kundu, P.~Bede{\l}ek, M.~Ostaszewski, O.~Danaci, Y.~J. Patel, V.~Dunjko, and J.~A. Miszczak, ``Enhancing variational quantum state diagonalization using reinforcement learning techniques,'' {\em New Journal of Physics}, vol.~26, no.~1, p.~013034, 2024.

\bibitem{Diag2}
A.~Sadhu, A.~Sarkar, and A.~Kundu, ``A quantum information theoretic analysis of reinforcement learning-assisted quantum architecture search,'' {\em Quantum Machine Intelligence}, vol.~6, no.~2, p.~49, 2024.

\bibitem{Diag3}
A.~Kundu, A.~Sarkar, and A.~Sadhu, ``Kanqas: Kolmogorov-arnold network for quantum architecture search,'' {\em EPJ Quantum Technology}, vol.~11, no.~1, p.~76, 2024.

\bibitem{Gumbel1954}
E.~Gumbel, ``Statistical theory of extreme values and some practical applications,'' {\em NBS Applied Mathematics Series}, 1954.

\bibitem{Bengio-gumbel-softmax16}
Y.~Bengio, N.~Leonard, and A.~Courville, ``Estimating or propagating gradients through stochastic neurons for conditional computation,'' {\em arXiv preprint arXiv:1308.3432}, 2013.

\bibitem{gumbel-softmax16}
E.~Jang, S.~Gu, and B.~Poole, ``Categorical reparameterization with gumbel-softmax,'' in {\em ICLR}, 2017.

\bibitem{sohl2015deep}
J.~Sohl-Dickstein, E.~Weiss, N.~Maheswaranathan, and S.~Ganguli, ``Deep unsupervised learning using nonequilibrium thermodynamics,'' in {\em ICML}, pp.~2256--2265, 2015.

\bibitem{rombach2022high}
R.~Rombach, A.~Blattmann, D.~Lorenz, P.~Esser, and B.~Ommer, ``High-resolution image synthesis with latent diffusion models,'' in {\em Proceedings of the IEEE/CVF conference on computer vision and pattern recognition}, pp.~10684--10695, 2022.

\bibitem{furrutter2024quantum}
F.~F{\"u}rrutter, G.~Mu{\~n}oz-Gil, and H.~J. Briegel, ``Quantum circuit synthesis with diffusion models,'' {\em Nature Machine Intelligence}, vol.~6, no.~5, pp.~515--524, 2024.

\bibitem{kliuchnikov2013optimization}
V.~Kliuchnikov and D.~Maslov, ``Optimization of clifford circuits,'' {\em Physical Review A}, vol.~88, no.~5, p.~052307, 2013.

\bibitem{xu2022quartz}
M.~Xu, Z.~Li, O.~Padon, S.~Lin, J.~Pointing, A.~Hirth, H.~Ma, J.~Palsberg, A.~Aiken, U.~A. Acar, {\em et~al.}, ``Quartz: superoptimization of quantum circuits,'' in {\em Proceedings of the 43rd ACM SIGPLAN International Conference on Programming Language Design and Implementation}, pp.~625--640, 2022.

\bibitem{pointing2021optimizing}
J.~Pointing, O.~Padon, Z.~Jia, H.~Ma, A.~Hirth, J.~Palsberg, and A.~Aiken, ``Quanto: Optimizing quantum circuits with automatic generation of circuit identities,'' {\em Quantum Science and Technology}, vol.~9, no.~4, p.~045009, 2024.

\bibitem{bravyi20226}
S.~Bravyi, J.~A. Latone, and D.~Maslov, ``6-qubit optimal clifford circuits,'' {\em npj Quantum Information}, vol.~8, no.~1, p.~79, 2022.

\bibitem{maslov2016advantages}
D.~Maslov, ``Advantages of using relative-phase toffoli gates with an application to multiple control toffoli optimization,'' {\em Physical Review A}, vol.~93, no.~2, p.~022311, 2016.

\bibitem{he2017decompositions}
Y.~He, M.-X. Luo, E.~Zhang, H.-K. Wang, and X.-F. Wang, ``Decompositions of n-qubit toffoli gates with linear circuit complexity,'' {\em International Journal of Theoretical Physics}, vol.~56, pp.~2350--2361, 2017.

\bibitem{bravyi2021clifford}
S.~Bravyi, R.~Shaydulin, S.~Hu, and D.~Maslov, ``Clifford circuit optimization with templates and symbolic pauli gates,'' {\em Quantum}, vol.~5, p.~580, 2021.

\bibitem{liu2021relaxed}
J.~Liu, L.~Bello, and H.~Zhou, ``Relaxed peephole optimization: A novel compiler optimization for quantum circuits,'' in {\em International Symposium on Code Generation and Optimization}, pp.~301--314, 2021.

\bibitem{sivarajah2020t}
S.~Sivarajah, S.~Dilkes, A.~Cowtan, W.~Simmons, A.~Edgington, and R.~Duncan, ``t$|\rm ket\rangle$: a retargetable compiler for nisq devices,'' {\em Quantum Science and Technology}, vol.~6, no.~1, p.~014003, 2020.

\bibitem{xu2023synthesizing}
A.~Xu, A.~Molavi, L.~Pick, S.~Tannu, and A.~Albarghouthi, ``Synthesizing quantum-circuit optimizers,'' {\em Proceedings of the ACM on Programming Languages}, vol.~7, no.~PLDI, pp.~835--859, 2023.

\bibitem{nam2020approximate}
Y.~Nam, Y.~Su, and D.~Maslov, ``Approximate quantum fourier transform with o (n log (n)) t gates,'' {\em NPJ Quantum Information}, vol.~6, no.~1, p.~26, 2020.

\bibitem{bae2020quantum}
J.-H. Bae, P.~M. Alsing, D.~Ahn, and W.~A. Miller, ``Quantum circuit optimization using quantum karnaugh map,'' {\em Scientific reports}, vol.~10, no.~1, p.~15651, 2020.

\bibitem{prasad2006data}
A.~K. Prasad, V.~V. Shende, I.~L. Markov, J.~P. Hayes, and K.~N. Patel, ``Data structures and algorithms for simplifying reversible circuits,'' {\em ACM Journal on Emerging Technologies in Computing Systems}, vol.~2, no.~4, pp.~277--293, 2006.

\bibitem{wu2020qgo}
X.-C. Wu, M.~G. Davis, F.~T. Chong, and C.~Iancu, ``Qgo: Scalable quantum circuit optimization using automated synthesis,'' {\em arXiv preprint arXiv:2012.09835}, 2020.

\bibitem{patel2021robust}
T.~Patel, E.~Younis, C.~Iancu, W.~de~Jong, and D.~Tiwari, ``Robust and resource-efficient quantum circuit approximation,'' {\em arXiv preprint arXiv:2108.12714}, 2021.

\bibitem{hietala2021verified}
K.~Hietala, R.~Rand, S.-H. Hung, X.~Wu, and M.~Hicks, ``A verified optimizer for quantum circuits,'' {\em Proceedings of the ACM on Programming Languages}, vol.~5, no.~POPL, pp.~1--29, 2021.

\bibitem{abdessaied2014quantum}
N.~Abdessaied, M.~Soeken, and R.~Drechsler, ``Quantum circuit optimization by hadamard gate reduction,'' in {\em Reversible Computation: 6th International Conference, RC 2014, Kyoto, Japan, July 10-11, 2014. Proceedings 6}, pp.~149--162, 2014.

\bibitem{zhang2019optimizing}
F.~Zhang and J.~Chen, ``Optimizing t gates in clifford+ t circuit as $\frac{\pi}{4}$ rotations around paulis,'' {\em arXiv preprint arXiv:1903.12456}, 2019.

\bibitem{iten2022exact}
R.~Iten, R.~Moyard, T.~Metger, D.~Sutter, and S.~Woerner, ``Exact and practical pattern matching for quantum circuit optimization,'' {\em ACM Transactions on Quantum Computing}, vol.~3, no.~1, pp.~1--41, 2022.

\bibitem{staudacher2023reducing}
K.~Staudacher, T.~Guggemos, S.~Grundner-Culemann, and W.~Gehrke, ``Reducing 2-qubit gate count for zx-calculus based quantum circuit optimization,'' {\em arXiv preprint arXiv:2311.08881}, 2023.

\bibitem{kissinger2020reducing}
A.~Kissinger and J.~van~de Wetering, ``Reducing the number of non-clifford gates in quantum circuits,'' {\em Physical Review A}, vol.~102, no.~2, p.~022406, 2020.

\bibitem{davis2019heuristics}
M.~G. Davis, E.~Smith, A.~Tudor, K.~Sen, I.~Siddiqi, and C.~Iancu, ``Heuristics for quantum compiling with a continuous gate set,'' {\em arXiv preprint arXiv:1912.02727}, 2019.

\bibitem{matsuzawa2020jastrow}
Y.~Matsuzawa and Y.~Kurashige, ``Jastrow-type decomposition in quantum chemistry for low-depth quantum circuits,'' {\em Journal of chemical theory and computation}, vol.~16, no.~2, pp.~944--952, 2020.

\bibitem{kattemolle2025edge}
J.~Kattem{\"o}lle, ``Edge coloring lattice graphs,'' {\em Journal of Mathematical Physics}, vol.~66, no.~5, 2025.

\bibitem{sun2023asymptotically}
X.~Sun, G.~Tian, S.~Yang, P.~Yuan, and S.~Zhang, ``Asymptotically optimal circuit depth for quantum state preparation and general unitary synthesis,'' {\em IEEE Transactions on Computer-Aided Design of Integrated Circuits and Systems}, vol.~42, no.~10, pp.~3301--3314, 2023.

\bibitem{yuan2023optimal}
P.~Yuan and S.~Zhang, ``Optimal (controlled) quantum state preparation and improved unitary synthesis by quantum circuits with any number of ancillary qubits,'' {\em Quantum}, vol.~7, p.~956, 2023.

\bibitem{yuan2023does}
P.~Yuan, J.~Allcock, and S.~Zhang, ``Does qubit connectivity impact quantum circuit complexity?,'' {\em IEEE Transactions on computer-aided design of integrated circuits and systems}, vol.~43, no.~2, pp.~520--533, 2023.

\bibitem{yuan2025full}
P.~Yuan and S.~Zhang, ``Full characterization of the depth overhead for quantum circuit compilation with arbitrary qubit connectivity constraint,'' {\em Quantum}, vol.~9, p.~1757, 2025.

\bibitem{herrman2021globally}
R.~Herrman, L.~Treffert, J.~Ostrowski, P.~C. Lotshaw, T.~S. Humble, and G.~Siopsis, ``Globally optimizing qaoa circuit depth for constrained optimization problems,'' {\em Algorithms}, vol.~14, no.~10, p.~294, 2021.

\bibitem{herrman2022multi}
R.~Herrman, P.~C. Lotshaw, J.~Ostrowski, T.~S. Humble, and G.~Siopsis, ``Multi-angle quantum approximate optimization algorithm,'' {\em Scientific Reports}, vol.~12, no.~1, p.~6781, 2022.

\bibitem{chandarana2022digitized}
P.~Chandarana, N.~N. Hegade, K.~Paul, F.~Albarr{\'a}n-Arriagada, E.~Solano, A.~Del~Campo, and X.~Chen, ``Digitized-counterdiabatic quantum approximate optimization algorithm,'' {\em Physical Review Research}, vol.~4, no.~1, p.~013141, 2022.

\bibitem{kivlichan2018quantum}
I.~D. Kivlichan, J.~McClean, N.~Wiebe, C.~Gidney, A.~Aspuru-Guzik, G.~K.-L. Chan, and R.~Babbush, ``Quantum simulation of electronic structure with linear depth and connectivity,'' {\em Physical review letters}, vol.~120, no.~11, p.~110501, 2018.

\bibitem{barenco1995elementary}
A.~Barenco, C.~H. Bennett, R.~Cleve, D.~P. DiVincenzo, N.~Margolus, P.~Shor, T.~Sleator, J.~A. Smolin, and H.~Weinfurter, ``Elementary gates for quantum computation,'' {\em Physical review A}, vol.~52, no.~5, p.~3457, 1995.

\bibitem{cybenko2001reducing}
G.~Cybenko, ``Reducing quantum computations to elementary unitary operations,'' {\em Computing in science \& engineering}, vol.~3, no.~2, pp.~27--32, 2001.

\bibitem{knill1995approximation}
E.~Knill, ``Approximation by quantum circuits,'' {\em arXiv preprint quant-ph/9508006}, 1995.

\bibitem{mottonen2004quantum}
M.~M{\"o}tt{\"o}nen, J.~J. Vartiainen, V.~Bergholm, and M.~M. Salomaa, ``Quantum circuits for general multiqubit gates,'' {\em Physical review letters}, vol.~93, no.~13, p.~130502, 2004.

\bibitem{shende2005synthesis}
V.~V. Shende, S.~S. Bullock, and I.~L. Markov, ``Synthesis of quantum logic circuits,'' in {\em Proceedings of the 2005 Asia and South Pacific Design Automation Conference}, pp.~272--275, 2005.

\bibitem{mottonen12006decompositions}
M.~M{\"o}tt{\"o}nen$^1$ and J.~J. Vartiainen, ``Decompositions of general quantum gates,'' {\em Trends in quantum computing research}, p.~149, 2006.

\bibitem{shende2008cnot}
V.~V. Shende and I.~L. Markov, ``On the cnot-cost of toffoli gates,'' {\em arXiv preprint arXiv:0803.2316}, 2008.

\bibitem{bullock2008asymptotically}
S.~S. Bullock and I.~L. Markov, ``Asymptotically optimal circuits for arbitrary n-qubit diagonal computations,'' {\em arXiv preprint quant-ph/0303039}, 2008.

\bibitem{jiang2020optimal}
J.~Jiang, X.~Sun, S.-H. Teng, B.~Wu, K.~Wu, and J.~Zhang, ``Optimal space-depth trade-off of cnot circuits in quantum logic synthesis,'' in {\em ACM-SIAM Symposium on Discrete Algorithms}, pp.~213--229, 2020.

\bibitem{gheorghiu2022reducing}
V.~Gheorghiu, J.~Huang, S.~M. Li, M.~Mosca, and P.~Mukhopadhyay, ``Reducing the cnot count for clifford+ t circuits on nisq architectures,'' {\em IEEE Transactions on Computer-Aided Design of Integrated Circuits and Systems}, 2022.

\bibitem{nash2020quantum}
B.~Nash, V.~Gheorghiu, and M.~Mosca, ``Quantum circuit optimizations for nisq architectures,'' {\em Quantum Science and Technology}, vol.~5, no.~2, p.~025010, 2020.

\bibitem{zhang2023low}
Z.-J. Zhang, J.~Sun, X.~Yuan, and M.-H. Yung, ``Low-depth hamiltonian simulation by an adaptive product formula,'' {\em Physical Review Letters}, vol.~130, no.~4, p.~040601, 2023.

\bibitem{yao2021adaptive}
Y.-X. Yao, N.~Gomes, F.~Zhang, C.-Z. Wang, K.-M. Ho, T.~Iadecola, and P.~P. Orth, ``Adaptive variational quantum dynamics simulations,'' {\em PRX Quantum}, vol.~2, no.~3, p.~030307, 2021.

\bibitem{motta2021low}
M.~Motta, E.~Ye, J.~R. McClean, Z.~Li, A.~J. Minnich, R.~Babbush, and G.~K.-L. Chan, ``Low rank representations for quantum simulation of electronic structure,'' {\em npj Quantum Information}, vol.~7, no.~1, p.~83, 2021.

\bibitem{tkachenko2021correlation}
N.~V. Tkachenko, J.~Sud, Y.~Zhang, S.~Tretiak, P.~M. Anisimov, A.~T. Arrasmith, P.~J. Coles, L.~Cincio, and P.~A. Dub, ``Correlation-informed permutation of qubits for reducing ansatz depth in the variational quantum eigensolver,'' {\em PRX Quantum}, vol.~2, no.~2, p.~020337, 2021.

\bibitem{chivilikhin2020mog}
D.~Chivilikhin, A.~Samarin, V.~Ulyantsev, I.~Iorsh, A.~Oganov, and O.~Kyriienko, ``Mog-vqe: Multiobjective genetic variational quantum eigensolver,'' {\em arXiv preprint arXiv:2007.04424}, 2020.

\bibitem{halder2024machine}
S.~Halder, A.~Dey, C.~Shrikhande, and R.~Maitra, ``Machine learning assisted construction of a shallow depth dynamic ansatz for noisy quantum hardware,'' {\em Chemical Science}, 2024.

\bibitem{anastasiou2024tetris}
P.~G. Anastasiou, Y.~Chen, N.~J. Mayhall, E.~Barnes, and S.~E. Economou, ``Tetris-adapt-vqe: An adaptive algorithm that yields shallower, denser circuit ans{\"a}tze,'' {\em Physical Review Research}, vol.~6, no.~1, p.~013254, 2024.

\bibitem{zhu2022adaptive}
L.~Zhu, H.~L. Tang, G.~S. Barron, F.~Calderon-Vargas, N.~J. Mayhall, E.~Barnes, and S.~E. Economou, ``Adaptive quantum approximate optimization algorithm for solving combinatorial problems on a quantum computer,'' {\em Physical Review Research}, vol.~4, no.~3, p.~033029, 2022.

\bibitem{zhou2000methodology}
X.~Zhou, D.~W. Leung, and I.~L. Chuang, ``Methodology for quantum logic gate construction,'' {\em Physical Review A}, vol.~62, no.~5, p.~052316, 2000.

\bibitem{aliferis2005quantum}
P.~Aliferis, D.~Gottesman, and J.~Preskill, ``Quantum accuracy threshold for concatenated distance-3 codes,'' {\em arXiv preprint quant-ph/0504218}, 2005.

\bibitem{fowler2009high}
A.~G. Fowler, A.~M. Stephens, and P.~Groszkowski, ``High-threshold universal quantum computation on the surface code,'' {\em Physical Review A}, vol.~80, no.~5, p.~052312, 2009.

\bibitem{bravyi2005universal}
S.~Bravyi and A.~Kitaev, ``Universal quantum computation with ideal clifford gates and noisy ancillas,'' {\em Physical Review A}, vol.~71, no.~2, p.~022316, 2005.

\bibitem{campbell2017roads}
E.~T. Campbell, B.~M. Terhal, and C.~Vuillot, ``Roads towards fault-tolerant universal quantum computation,'' {\em Nature}, vol.~549, no.~7671, pp.~172--179, 2017.

\bibitem{davis2020towards}
M.~G. Davis, E.~Smith, A.~Tudor, K.~Sen, I.~Siddiqi, and C.~Iancu, ``Towards optimal topology aware quantum circuit synthesis,'' in {\em 2020 IEEE International Conference on Quantum Computing and Engineering (QCE)}, pp.~223--234, 2020.

\bibitem{aaronson2004improved}
S.~Aaronson and D.~Gottesman, ``Improved simulation of stabilizer circuits,'' {\em Physical Review A}, vol.~70, no.~5, p.~052328, 2004.

\bibitem{fawzi2022discovering}
A.~Fawzi, M.~Balog, A.~Huang, T.~Hubert, B.~Romera-Paredes, M.~Barekatain, A.~Novikov, F.~J. R~Ruiz, J.~Schrittwieser, G.~Swirszcz, {\em et~al.}, ``Discovering faster matrix multiplication algorithms with reinforcement learning,'' {\em Nature}, vol.~610, no.~7930, pp.~47--53, 2022.

\bibitem{tilly2022variational}
J.~Tilly, H.~Chen, S.~Cao, D.~Picozzi, K.~Setia, Y.~Li, E.~Grant, L.~Wossnig, I.~Rungger, G.~H. Booth, {\em et~al.}, ``The variational quantum eigensolver: a review of methods and best practices,'' {\em Physics Reports}, vol.~986, pp.~1--128, 2022.

\bibitem{fedorov2022vqe}
D.~A. Fedorov, B.~Peng, N.~Govind, and Y.~Alexeev, ``Vqe method: a short survey and recent developments,'' {\em Materials Theory}, vol.~6, no.~1, p.~2, 2022.

\bibitem{cerezo2021variational}
M.~Cerezo, A.~Arrasmith, R.~Babbush, S.~C. Benjamin, S.~Endo, K.~Fujii, J.~R. McClean, K.~Mitarai, X.~Yuan, L.~Cincio, {\em et~al.}, ``Variational quantum algorithms,'' {\em Nature Reviews Physics}, vol.~3, no.~9, pp.~625--644, 2021.

\bibitem{blekos2024review}
K.~Blekos, D.~Brand, A.~Ceschini, C.-H. Chou, R.-H. Li, K.~Pandya, and A.~Summer, ``A review on quantum approximate optimization algorithm and its variants,'' {\em Physics Reports}, vol.~1068, pp.~1--66, 2024.

\bibitem{kandala2017hardware}
A.~Kandala, A.~Mezzacapo, K.~Temme, M.~Takita, M.~Brink, J.~M. Chow, and J.~M. Gambetta, ``Hardware-efficient variational quantum eigensolver for small molecules and quantum magnets,'' {\em nature}, vol.~549, no.~7671, pp.~242--246, 2017.

\bibitem{romero1701strategies}
J.~Romero, R.~Babbush, J.~R. McClean, C.~Hempel, P.~J. Love, and A.~Aspuru-Guzik, ``Strategies for quantum computing molecular energies using the unitary coupled cluster ansatz,'' {\em Quantum Science and Technology}, vol.~4, no.~1, p.~014008, 2018.

\bibitem{burton2023exact}
H.~G. Burton, D.~Marti-Dafcik, D.~P. Tew, and D.~J. Wales, ``Exact electronic states with shallow quantum circuits from global optimisation,'' {\em npj Quantum Information}, vol.~9, no.~1, p.~75, 2023.

\bibitem{hinton2012practical}
G.~E. Hinton, ``A practical guide to training restricted boltzmann machines,'' in {\em Neural Networks: Tricks of the Trade: Second Edition}, pp.~599--619, Springer, 2012.

\bibitem{alam2020circuit}
M.~Alam, A.~Ash-Saki, and S.~Ghosh, ``Circuit compilation methodologies for quantum approximate optimization algorithm,'' in {\em IEEE/ACM MICRO}, pp.~215--228, 2020.

\bibitem{arufe2022quantum}
L.~Arufe, M.~A. Gonz{\'a}lez, A.~Oddi, R.~Rasconi, and R.~Varela, ``Quantum circuit compilation by genetic algorithm for quantum approximate optimization algorithm applied to maxcut problem,'' {\em Swarm and Evolutionary Computation}, vol.~69, p.~101030, 2022.

\bibitem{west2001introduction}
D.~B. West {\em et~al.}, {\em Introduction to graph theory}, vol.~2.
\newblock Prentice hall Upper Saddle River, 2001.

\bibitem{vizing1964estimate}
V.~G. Vizing, ``On an estimate of the chromatic class of a p-graph,'' {\em Diskret analiz}, vol.~3, pp.~25--30, 1964.

\bibitem{yanakiev2024dynamic}
N.~Yanakiev, N.~Mertig, C.~K. Long, and D.~R. Arvidsson-Shukur, ``Dynamic adaptive quantum approximate optimization algorithm for shallow, noise-resilient circuits,'' {\em Physical Review A}, vol.~109, no.~3, p.~032420, 2024.

\bibitem{XQAOA}
V.~Vijendran, A.~Das, D.~E. Koh, S.~M. Assad, and P.~K. Lam, ``An expressive ansatz for low-depth quantum approximate optimisation,'' {\em Quantum Science and Technology}, vol.~9, no.~2, p.~025010, 2024.

\bibitem{cowtan2019qubit}
A.~Cowtan, S.~Dilkes, R.~Duncan, A.~Krajenbrink, W.~Simmons, and S.~Sivarajah, ``On the qubit routing problem,'' {\em arXiv preprint arXiv:1902.08091}, 2019.

\bibitem{zhu2020dynamic}
P.~Zhu, Z.~Guan, and X.~Cheng, ``A dynamic look-ahead heuristic for the qubit mapping problem of nisq computers,'' {\em IEEE Transactions on Computer-Aided Design of Integrated Circuits and Systems}, vol.~39, no.~12, pp.~4721--4735, 2020.

\bibitem{li2020qubit}
S.~Li, X.~Zhou, and Y.~Feng, ``Qubit mapping based on subgraph isomorphism and filtered depth-limited search,'' {\em IEEE Transactions on Computers}, vol.~70, no.~11, pp.~1777--1788, 2020.

\bibitem{zhou2020quantum}
X.~Zhou, S.~Li, and Y.~Feng, ``Quantum circuit transformation based on simulated annealing and heuristic search,'' {\em IEEE Transactions on Computer-Aided Design of Integrated Circuits and Systems}, vol.~39, no.~12, pp.~4683--4694, 2020.

\bibitem{zhou2020monte}
X.~Zhou, Y.~Feng, and S.~Li, ``A monte carlo tree search framework for quantum circuit transformation,'' in {\em ICCAD}, pp.~1--7, 2020.

\bibitem{park2022fast}
S.~Park, D.~Kim, M.~Kweon, J.-Y. Sim, and S.~Kang, ``A fast and scalable qubit-mapping method for noisy intermediate-scale quantum computers,'' in {\em DAC}, pp.~13--18, 2022.

\bibitem{nannicini2022optimal}
G.~Nannicini, L.~S. Bishop, O.~G{\"u}nl{\"u}k, and P.~Jurcevic, ``Optimal qubit assignment and routing via integer programming,'' {\em ACM Transactions on Quantum Computing}, vol.~4, no.~1, pp.~1--31, 2022.

\bibitem{zhang2021time}
C.~Zhang, A.~B. Hayes, L.~Qiu, Y.~Jin, Y.~Chen, and E.~Z. Zhang, ``Time-optimal qubit mapping,'' in {\em ACM International Conference on Architectural Support for Programming Languages and Operating Systems}, pp.~360--374, 2021.

\bibitem{ash2019qure}
A.~Ash-Saki, M.~Alam, and S.~Ghosh, ``Qure: Qubit re-allocation in noisy intermediate-scale quantum computers,'' in {\em Annual Design Automation Conference}, pp.~1--6, 2019.

\bibitem{das2019case}
P.~Das, S.~S. Tannu, P.~J. Nair, and M.~Qureshi, ``A case for multi-programming quantum computers,'' in {\em IEEE/ACM International Symposium on Microarchitecture}, pp.~291--303, 2019.

\bibitem{wu2024reducing}
W.~Wu, Y.~Wang, G.~Yan, Y.~Zhao, B.~Zhang, and J.~Yan, ``On reducing the execution latency of superconducting quantum processors via quantum job scheduling,'' in {\em ICCAD}, pp.~1--9, 2024.

\bibitem{huang2017homomorphic}
H.-L. Huang, Y.-W. Zhao, T.~Li, F.-G. Li, Y.-T. Du, X.-Q. Fu, S.~Zhang, X.~Wang, and W.-S. Bao, ``Homomorphic encryption experiments on ibm’s cloud quantum computing platform,'' {\em Frontiers of Physics}, vol.~12, pp.~1--6, 2017.

\bibitem{huang2018demonstration}
H.-L. Huang, A.~K. Goswami, W.-S. Bao, and P.~K. Panigrahi, ``Demonstration of essentiality of entanglement in a deutsch-like quantum algorithm,'' {\em Science China Physics, Mechanics \& Astronomy}, vol.~61, pp.~1--7, 2018.

\bibitem{caleffi2024distributed}
M.~Caleffi, M.~Amoretti, D.~Ferrari, J.~Illiano, A.~Manzalini, and A.~S. Cacciapuoti, ``Distributed quantum computing: a survey,'' {\em Computer Networks}, vol.~254, p.~110672, 2024.

\end{thebibliography}
\end{document}